\documentclass[
nofootinbib,
reprint,
superscriptaddress,
aps
]{revtex4-2}
\usepackage{physics}
\usepackage{bm}
\usepackage[T1]{fontenc}
\usepackage{amsmath,amsfonts,bbm, graphicx, color}
\usepackage{amssymb}
\usepackage{appendix}
\usepackage[colorlinks]{hyperref}
\usepackage[figure,table]{hypcap}
\usepackage{mathtools}

\usepackage{relsize}
\usepackage{xcolor}
\usepackage{scalerel}
\usepackage{tikz}
\usetikzlibrary{svg.path}
\usepackage{siunitx}

\definecolor{orcidlogocol}{HTML}{A6CE39}
\tikzset{
  orcidlogo/.pic={
    \fill[orcidlogocol] svg{M256,128c0,70.7-57.3,128-128,128C57.3,256,0,198.7,0,128C0,57.3,57.3,0,128,0C198.7,0,256,57.3,256,128z};
    \fill[white] svg{M86.3,186.2H70.9V79.1h15.4v48.4V186.2z}
                 svg{M108.9,79.1h41.6c39.6,0,57,28.3,57,53.6c0,27.5-21.5,53.6-56.8,53.6h-41.8V79.1z M124.3,172.4h24.5c34.9,0,42.9-26.5,42.9-39.7c0-21.5-13.7-39.7-43.7-39.7h-23.7V172.4z}
                 svg{M88.7,56.8c0,5.5-4.5,10.1-10.1,10.1c-5.6,0-10.1-4.6-10.1-10.1c0-5.6,4.5-10.1,10.1-10.1C84.2,46.7,88.7,51.3,88.7,56.8z};
  }
}

\newcommand\orcid[1]{\href{https://orcid.org/#1}{\mbox{\scalerel*{
\begin{tikzpicture}[yscale=-1,transform shape]
\pic{orcidlogo};
\end{tikzpicture}
}{|}}}}
\usepackage[colorlinks]{hyperref}

\hypersetup{
	bookmarksnumbered,
	pdfstartview={FitH},
	citecolor={blue},
	linkcolor={blue},
	urlcolor={blue},
	pdfpagemode={UseOutlines}}
\definecolor{darkgreen}{RGB}{20,100,20}
\definecolor{darkblue}{RGB}{0,0,130}
\definecolor{darkred}{rgb}{.8,0,0}

\usepackage{soul}

\begin{document}

\title{Single- to many-body crossover of a quantum carpet}

\author{Maciej \L{}ebek\orcid{0000-0003-4858-2460}}
\affiliation{Center for Theoretical Physics, Polish Academy of Sciences, Aleja Lotnik\'ow 32/46, 02-668 Warsaw, Poland}
\affiliation{Institute of Theoretical Physics, University of Warsaw, Pasteura 5, 02-093 Warsaw, Poland}

\author{Piotr T. Grochowski\orcid{0000-0002-9654-4824}}
\email{piotr@cft.edu.pl}
\affiliation{Center for Theoretical Physics, Polish Academy of Sciences, Aleja Lotnik\'ow 32/46, 02-668 Warsaw, Poland}

\author{Kazimierz Rz\k{a}\.zewski\orcid{0000-0002-6082-3565}}
\affiliation{Center for Theoretical Physics, Polish Academy of Sciences, Aleja Lotnik\'ow 32/46, 02-668 Warsaw, Poland}
\date{\today}

\begin{abstract}
Strongly interacting many-body system of bosons exhibiting the quantum carpet pattern is investigated exactly by using Gaudin solutions. 
We show that this highly coherent design usually present in noninteracting, single-body scenarios gets destroyed by weak-to-moderate interatomic interactions in an ultracold bosonic gas trapped in a box potential.
However, it becomes revived in a very strongly interacting regime, when the system undergoes fermionization.
We track the whole single- to many-body crossover, providing an analysis of de- and rephasing present in the system.

\end{abstract}
\maketitle
\section{Introduction}
Periodic self-replication of physical systems have been known since at least nineteenth century when Henry Fox Talbot discovered spatially repeating patterns in his experiments with diffraction gratings~\cite{Talbot1836,LordRayleigh1881}.
Such recurrences are most often associated with underlying wave nature of the system with light being the most straightforward example.
As such, it comes as a no surprise that also quantum mechanical objects like ultracold atoms exhibit similar self-repeating behavior, however also in a time domain, giving rise to a rich family of phenomena, among others, quantum fractals~\cite{Berry1996,Wojcik2000,Gao2019}, quantum echoes~\cite{Buchkremer2000}, quantum Talbot effect~\cite{Sanz2007}, and quantum scars~\cite{Heller1984,Kaplan1998}, collectively addressed as quantum revivals~\cite{Eberly1980,Robinett2004}.

\begin{figure*}[t]
	\includegraphics[width=1.0\linewidth]{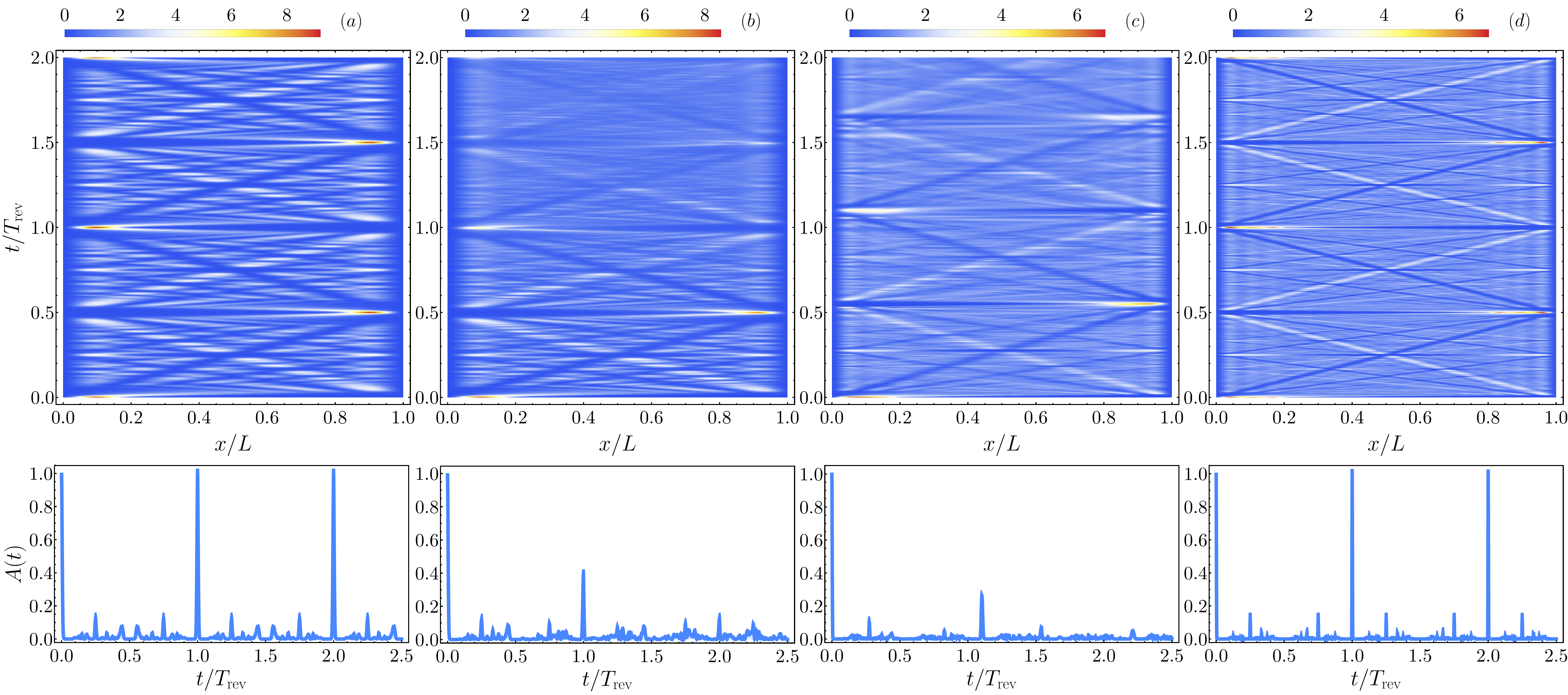}
	\caption{\label{fig1}Top row: Quantum carpets of $N=3$ bosons which are ideal (a), interacting repulsively with $\gamma=0.86$ (b), $\gamma=20.0$ (c) or fermionized (d).
	The time coherence is preserved in the first and the last cases, while a finite interaction smears out the pattern, signalling loss of coherence in the system.
	Bottom row: The squares of the autocorrelation functions for each of these cases.
	Perfect revivals are present at the multiples of the revival time, $T_{\text{rev}}$, for ideal and fermionized bosons.
	Imperfect revivals are visible for intermediate interaction strengths, being shifted away from $T_{\text{rev}}$ (see panel (c)).}
\end{figure*}

One of the most aesthetically appealing examples of such revival phenomena are quantum carpets---spatiotemporal depictions of probability density of an initially localized quantum particle in a box potential (see Fig.~\ref{fig1}(a))~\cite{Kinzel1995,Stifter1997,Grossmann1997,Marzoli1998,Kaplan1998a,Berry2001}.
Their distinctive features are called \textit{canals} and \textit{ridges}---characteristic lines minimizing or maximizing the probability density.
These structures have been proposed to act as decoherence probes, giving direct access to intermode coherence at the level of a single particle density~\cite{Kazemi2013,Chen2018}.
Despite being studied by various approaches, including Wigner representation~\cite{Marzoli1998,Friesch2000}, degeneracy in intermode traces~\cite{Grossmann1997,Stifter1997,Marzoli1998,Friesch2000,Loinaz1999}, travelling wave decomposition~\cite{Hall1999}, spin chains~\cite{Banchi2015,Genest2016a,Genest2016,Lemay2016,Kay2017,Compagno2016,Christandl2017}, and fractional revivals~\cite{Aronstein1997}, their presence in many-body interacting systems was rather insufficiently examined~\cite{Nest2006}.
These scarce investigations were based either on mean-field models in bosonic systems~\cite{Ruostekoski2001,Gawryluk2006} or weakly interacting fermionic ones~\cite{Nest2006,Grochowski2020}---the full many-body analysis has been absent in the literature.

However, recent study showed that even in the simplest many-body system of ideal fermions interesting phenomena arise~\cite{Grochowski2020}.
The quantum carpet design gets much more pronounced, with canals and ridges becoming solitonlike---narrower and more distinct (see Fig.~\ref{fig1}(d)). 
Such a fermionic quantum carpet appears when an ultracold fermionic gas is initially trapped in a box and then instantaneously released to a bigger one.
Such a scenario is necessary to sustain a quantum carpet design in the large particle number limit, as other initial trappings cause destruction of a regular pattern.
This strategy of preparing the nonequilibrium state might strike as troublesome, however experimental techniques involving atomic diffraction gratings or phase imprinting may reproduce similar dynamics~\cite{Nowak1997,Chapman1995,Ryu2006,Deng1999,Mark2011,Ahn2001,Katsuki2009}.

In our work, we show  that at the level of a single-particle density, a gas of strongly interacting bosons in the limit of fermionization retrieves the same fermionic carpet behavior.
We stress the distinction between this strongly correlated case and prototypical one of ideal fermions, as the differences appear already for a reduced single-particle density matrix, uncovering underlying structures of correlations and coherence.
Moreover, as a noninteracting bosonic cloud exhibits a typical single-particle quantum carpet pattern, for a finite interaction strength between atoms there is some nontrivial crossover between a single- and many-body carpet of fermionized particles.

We perform the analysis of this crossover basing on Gaudin solutions from the Bethe Ansatz~\cite{Bethe1931,Lieb1963b,Gaudin1971,Batchelor2005,Oelkers2006a,Gaudin2012,Tomchenko2015}, which allows us to have access to exact dynamics after arbitrary long evolution time, contrary to previous approaches based on approximate methods.
We utilize this advantage to investigate short and long time intrinsic dephasing in the system caused by introduction of interaction and eventual rephasing that happens in the nearly fermionized regime.
Along the lines of previous studies~\cite{Kazemi2013}, we provide an explicit physical mechanism in which quantum carpets give access to coherence and show how the de- and rephasing can be probed for highly nonequilibrium many-body states.

\section{Model}
Dynamics of highly nonequilibrium and strongly interacting systems, such as quantum carpets, is demanding to treat exactly~\cite{Syrwid2016,Staron2020,Syrwid2020}.
One of the examples of the system that is integrable and allows for an exact treatment is the Lieb-Liniger model of $N$ bosons interacting by a contact delta potential:
\begin{equation}
    H=-\frac{\hbar^2}{2 m}\sum_{i=1}^N \frac{\partial^2}{\partial x_i^2}+\frac{N \hbar^2}{m L} \gamma \sum_{i<j}\delta(x_i-x_j)
\end{equation}
where $x_1,\ldots,x_N$ are positions of atoms, $m$ is their mass, $L$ is a length of the box, $\gamma$ is a dimensionless parameter quantifying interaction strength and is connected to Lieb-Liniger coupling constant $c$ by $\frac{ N \gamma  }{ L} =  c$.
Parameter $\gamma$ is positive since we consider repulsive interactions only.
We are interested in the less explored case of hard-wall boundary conditions solved exactly by M. Gaudin~\cite{Gaudin1971,Gaudin2012}.
The eigenstates of the system take the form of complex superpositions of plane waves
\begin{equation}
\label{bosesym}
\begin{aligned}
    \psi_{\{k\}}&(\{x\})= \frac{1}{\sqrt{\mathcal{N}_{\{k\}}}} \sum_{\{\epsilon\}}\sum_P \epsilon_1 \ldots \epsilon_N \times \\ 
    &\prod_{i<j}\bigg(1-\frac{ic}{\epsilon_i k_i+ \epsilon_j k_j}\bigg)
    \bigg(1+\frac{ic \, \text{sgn}(x_j-x_i)}{\epsilon_{Pi}k_{Pi}-\epsilon_{Pj}k_{Pj}} \bigg)\times \\
    & \exp \big[i(\epsilon_{P1} k_{P1}x_1+\ldots+\epsilon_{PN} k_{PN}x_N)\big].
\end{aligned}
\end{equation}
By $\{x\}$ we denote $x_1,\ldots,x_N$, integers $\epsilon_i,  \, i=1,\ldots,N$ take two values $\pm 1$ and summing over $\{\epsilon \}$ means sum over all $2^N$ possibilities $\epsilon_1,\ldots,\epsilon_N$. The sum over $P$ runs through all permutations $P \in S_N$ and constants $\mathcal{N}_{\{k\}}$ assure proper normalization. The states are parametrized by sets of $N$ positive numbers $\{k\}$ called quasi-momenta obtained by solving equations
\begin{equation}
\label{betheEqs}
    k_i L = \pi n_i +\sum_{j \neq i} \Bigg( \arctan \frac{c}{k_i-k_j} +\arctan \frac{c}{k_i+k_j} \Bigg),
\end{equation}
where $i=1,\ldots,N$.
Set $\{k\}$ directly corresponds to the set $\{n\}, \, 1 \leq n_1\leq\ldots\leq n_N$, which can be understood as a set of quantum numbers for our system.
\begin{figure*}[t]
	\includegraphics[width=1.0\linewidth]{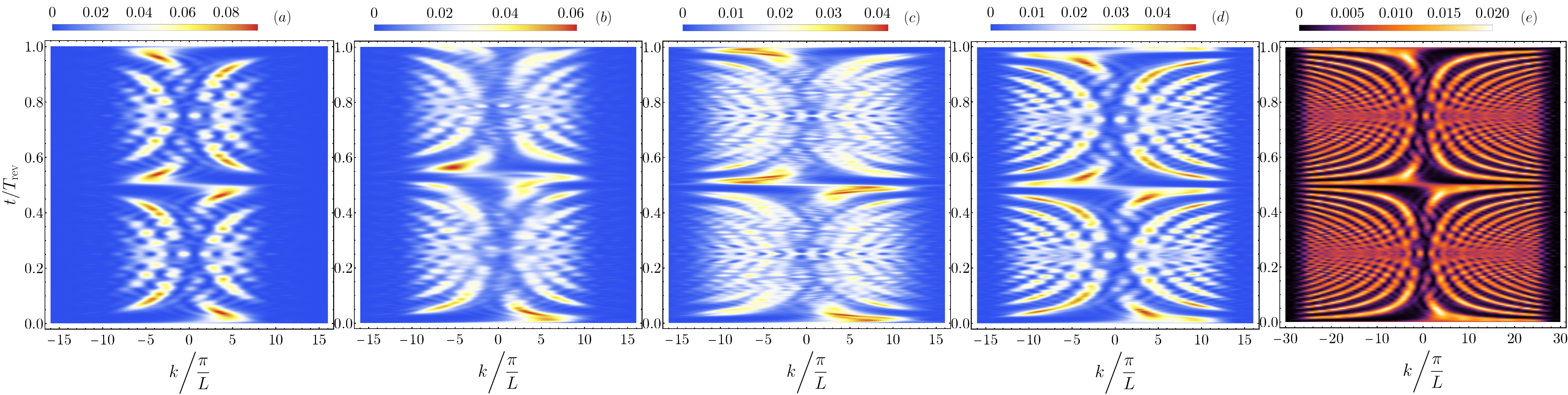}
	\caption{\label{fig2}Momentum quantum carpets of $N=2$ bosons which are ideal (a), interacting repulsively with $\gamma=5.8$ (b) or fermionized (c). 
	Two last carpets present an evolution of the momentum distribution for $N=2$ (d) and $N=5$ (e) ideal fermions.
	Note that there is a visible difference between fermionized bosons and ideal fermions, contrary to quantum carpet in the position space.
	Once again the loss of temporal coherence due to finite interactions is apparent (b).}
\end{figure*}
Eigenenergies are equal to $E_{\{k\}}=\sum_{i=1}^N\frac{\hbar^2 k_i^2}{2m}$.

Throughout this work, we will consider a very particular scenario of exciting the many-body quantum system.
Let the quantum gas of $N$ atoms be initially confined inside a box potential with the length of $D < L$ and stay in its ground state (throughout the work we consider a particular value $D=0.21 L$).
Then, it gets released into a bigger box with the length of $L$ that shares one of its walls with the initial one.
In order to obtain time-evolved wave function one has to solve equations and calculate overlaps $C_{\{k\}}$ between the initial state and eigenstates of the box with the width $L$.
This can be done analytically and we present details of that calculation in the Appendix~\ref{appb}.
In practice it is sufficient to take finite number $M$ of eigenstates in order to represent the initial state accurately.
Thus, time-evolved wave function reads
\begin{equation}
    \Psi(\{x\},t)=\sum_{u=1}^M e^{-iE_ut}C_u \psi_u(\{x\}).
\end{equation}

Quantum carpets are visible in the evolution of the diagonal part ($x=x'$) of reduced single-particle density matrix
\begin{equation}
\label{single}
    \rho(x,x',t)= \int \dd x_N  \ldots \int \dd x_2  \Psi^*\left(x,\ldots,t\right) \Psi\left(x',\ldots,t\right).
\end{equation}
It is possible to calculate single-particle density matrix analytically, details are presented in the Appendix~\ref{appb}.

While the diagonal part of~\eqref{single} describes the distribution of particles in the position space, in principle one can also look at the momentum space equivalent

\begin{equation}
    \tilde{\rho}(k,k',t)=\frac{1}{2\pi}\int \dd x \dd x' \rho(x,x',t) e^{-ikx} e^{ik'x'}.
\end{equation}
We will consider the diagonal part $k=k'$ only---single-particle density in the momentum space.
From the experimental point of view, access to the momentum distribution can be obtained through time-of-flight measurements, also for quasi-one-dimensional gases as considered in here~\cite{Wilson2020}.

While complexity of the expressions involved allowed us to obtain results for $N=2, 3$ particles\footnote{The number of terms in exact solution scales as $2^N N!$, which is faster than the usual scaling of Lieb-Liniger eigenstates.}, the evaluation of density matrices becomes much easier in the case of ideal bosons, ideal fermions and fermionized bosons i.e. in the limit $c \to \infty$. 
\begin{figure*}[t]
	\includegraphics[width=1.\linewidth]{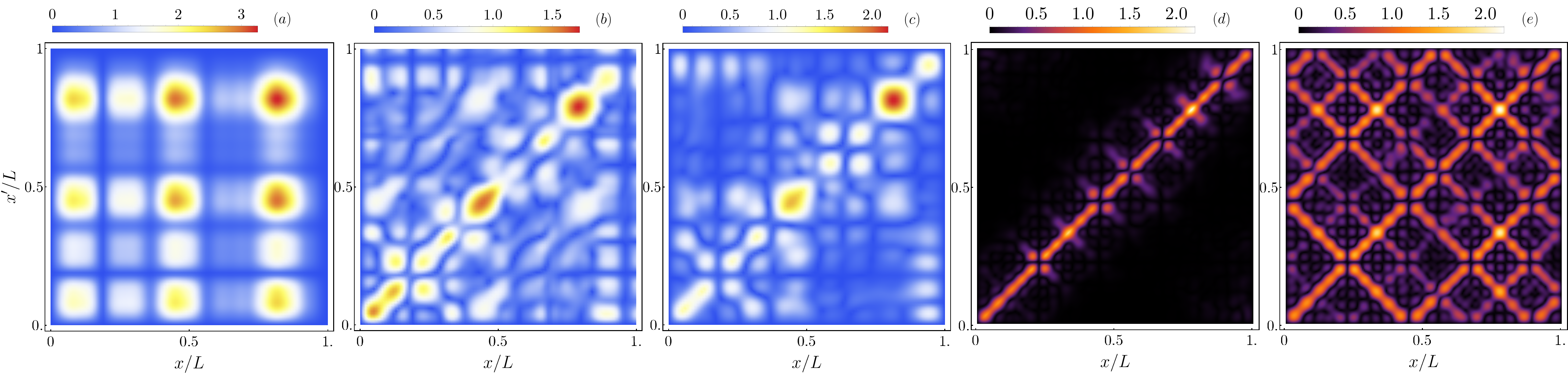}
	\caption{\label{fig3}Modulus of reduced single-particle density matrices for $N=2$ bosons that are ideal (a), interacting repulsively with $\gamma=22.5$ (b), fermionized (c) at the time $t=0.0914 T_{\text{rev}}$ and for $N=5$ fermionized bosons (d) or ideal fermions (e) at the time $t=0.0725 T_{\text{rev}}$.
	The spatial coherence is present all across the system in the ideal bosonic and ideal fermionic cases, however the structure of coherence is visibly different---in the latter case rectangular-shaped, narrow structures are present signifying correlations between solitonlike canals and ridges from the fermionic quantum carpet.
	Fermionized bosons do not recreate this type of coherence, with off-diagonal parts decaying as one goes away from the diagonal.}
\end{figure*}

\section{Quantum carpets}
Let us start with ideal fermions, which are not described by the model just introduced, but generate a prototypical fermionic quantum carpet~\cite{Grochowski2020}.
The wave function of ideal gas of fermions is given by the single Slater determinant: $\Psi_f (x_1,...,x_{N}) = \frac{1}{\sqrt{N!}} \det \left( \phi_1, \cdots,\phi_N \right)$, where $\phi_i(x),\, {i=1,...,N}$ denote different, orthonormal orbitals. 
When released into a box with the length of $L$, each of the orbitals starts to evolve unitarily,  $\phi_n(x,t)=\sum_{k=1}^{\infty} \Lambda_{n,k} \varphi_k (x) \exp\left( {-i E_k t / \hbar}\right) $.
Here, $\varphi_k(x) = \sqrt{2/L} \sin{\left( k \pi x/L  \right) } \theta{\left( x \right)} \theta{ \left( L- x \right) }, k=1,2,\ldots,$ are mode solutions of a large box, where $\theta(x)$ is a Heaviside step function and eigenenergies read $E_k=k^2 \pi^2 \hbar^2/2 m L^2$.
$\Lambda_{n,k} \equiv \bra{ \varphi_k}\ket{\phi_n }$ are overlaps between initial orbitals $\phi_n$ and box trap eigenfunctions $\varphi_k$.

We contrast this case to the fermionization regime, when bosons interact repulsively with an infinite strength.
There, the wave function is given by a Tonks-Girardeau form: $\Psi_{bf} (x_1,...,x_{N}) = \frac{1}{\sqrt{N!}} \prod_{1\leq i < j \leq N} \text{sgn} \left( x_i - x_j \right) \det \left( \phi_1, \cdots,\phi_N \right)$~\cite{Girardeau1960a,Yukalov2005}.
The Slater determinant is preceded by an unit antisymmetric function guaranteeing correct bosonic symmetry of the whole wave function.
The orbitals and their evolution are the same in both cases and so is single-particle density, $\rho (x,t) = \sum_{n=1}^{N} \left| \phi_n (x,t) \right|^2$.

When looking at the single particle density of this quantum gas in a spatiotemporal plot, an excitation scheme we consider produces a very thoroughly studied single-body quantum carpet pattern if the gas consists of ideal bosons (Fig.~\ref{fig1}(a)).
On the other hand, if ideal fermions are considered, the pattern changes, getting more pronounced, yielding so-called fermionic quantum carpet (Fig.~\ref{fig1}(d)).
It is necessary to note that such a behavior in the fermionic and fermionized cases is specific for this type of excitation---should the gas be initially confined in e.g. harmonic trap, the regularity would be lost~\cite{Grochowski2020}.

With the help of the Bethe Ansatz, we can also take a look at the quantum carpet design (see Fig.~\ref{fig1}) for the interacting system.
Not surprisingly, the weak interaction destroys the ideal single-particle quantum carpet, blurring the plot after a long evolution time.
However, the regularity becomes revived as the system becomes strongly interacting, eventually arriving at the fermionic quantum carpet in the case of fermionization.

Routinely, the difference between ideal fermions and fermionized bosons is extracted from their momentum distributions---the distribution for fermions is broader when the ground state is considered.
For fermionized bosons, the analysis of their momentum distribution's time evolution after a geometric quench in a harmonic trap has been analyzed both theoretically~\cite{Rigol2005,Minguzzi2005} and experimentally~\cite{Wilson2020}, showing signatures of so-called dynamical fermionization---broadening of the distribution during the evolution.
We also show such an analysis in Fig.~\ref{fig2} for interacting bosons and ideal fermions, arriving at \textit{momentum quantum carpets}.
Note that for ideal fermions an analytic expression can be derived, allowing us to evaluate the carpet for a larger number of atoms (see Fig.~\ref{fig2}(e)).
For all analyzed cases we find curved, arc-like structures of pronounced and reduced probability density, similar to ridges and canals from the spatiotemporal picture.
Initially well localized, the momentum distribution gets broader during the evolution, nonetheless revivng at $T_{\text{rev}}$ for ideal and fermionized cases.
Again, for a finite interaction, these structures get blurred with revivals becoming imperfect.

For ideal fermions, the reduced single-particle density matrix can be expressed as
\begin{align} \label{eq1}
\rho_f \left( x,x',t\right) = &\sum ^{N }_{n=1}\sum _{\pm }\sum _{p\in \mathbb{Z} }\sum ^{\infty }_{k=1} \pm \frac{1}{L}\Lambda_{n,k}\Lambda_ {n,k+ \left| p\right|} \times \\ \nonumber
 &  e^{i \frac{\pi}{L} p \frac{x' \pm x}{2}} \cos \left[\frac{\pi }{L}\left( \left| p\right| +2k\right)  \left( \frac{x' \mp x}{2}-pv_{0}t\right) \right],
\end{align}
where $v_0 = \pi \hbar/2 m L$ is the characteristic velocity that is connected to time of the system's revival $T_{\text{rev}}=2 L/ v_0 = 4 L^2 m/\pi \hbar$.
The formula~\eqref{eq1} is derived in the Appendix~\ref{appa}.

The density matrix can be expressed as a sum of travelling contributions that move with the velocities that are multiples of $v_0$, where $p$ denotes each of these terms.
Each travelling contribution is associated with rectangular-shaped breathing structure in the density matrix, whose shape is roughly described by $\left({x' \mp x}\right) /2-pv_{0}t = 0$ (see Fig.~\ref{fig3}(e)).
These structures are connected to the solitonlike objects~\cite{Grochowski2020} from the single-particle density and their presence is a signal of a coherence in the system.

The time evolved density matrix in case of fermionized bosons cannot be cast into a simple form of Eq.~\eqref{eq1}.
However, its fast numerical evaluation is possible, which we take advantage of~\cite{Pezer2007a}.
The rectangular-shaped structures are suppressed as the off-diagonal terms decay the faster the further they are from the diagonal, which is a characteristic feature of a Tonks-Girardeau gas.
However, temporal coherence is preserved as the system revives exactly at the multiples of the revival time, $T_{\text{rev}}$.
Note the revival time is the same for ideal bosons, ideal fermions and fermionized bosons. 

In case of bosons with a finite interaction strength (Fig~\ref{fig3}(a)-(c)), the density matrix is explicitly spatially coherent for a noninteracting system and its coherence length becomes visibly shorter for growing interaction, achieving its fermionized value for an infinite repulsion.
It shows that while the temporal coherence is restored, the spatial one is lost, submitting to the general feature of a fermionized system.

\begin{figure}[!ht]
	\includegraphics[width=1.\linewidth]{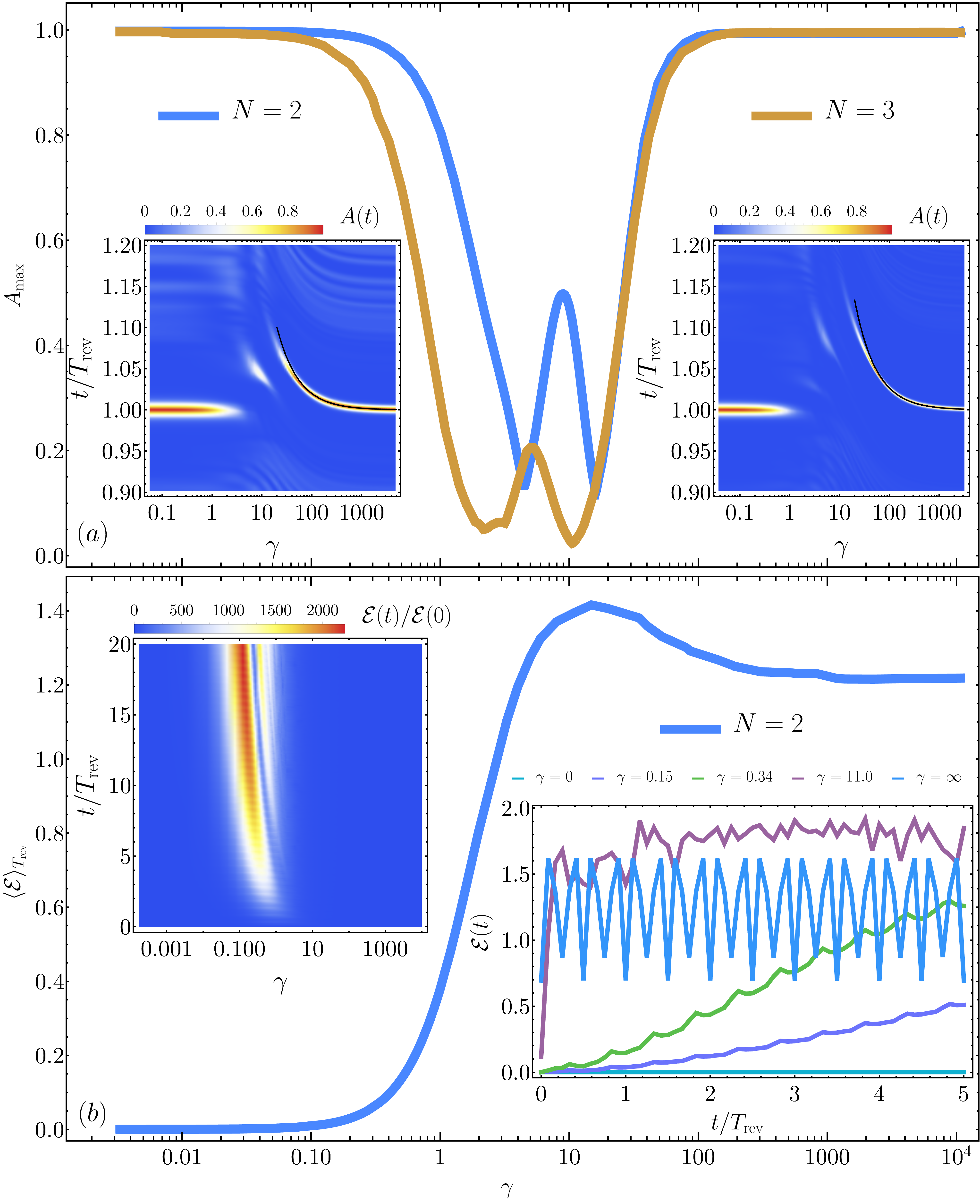}
	\caption{\label{fig4}(a) Maximum of a square of an autocorrelation function, $A_{\text{max}}$, close to the revival time $T_{\text{rev}}$ for $N=2$ and $N=3$ bosons for different interaction strengths.
	The decrease of the value signifies the imperfect revival.
	Left (2 atoms) and right (3 atoms) insets show a density plot of $A(t)$ close to $T_{\text{rev}}$.
	One can see that the revival time does not change for a weak interaction, while it increases with diminishing of interaction strength close to fermionization.
	Additionally, for intermediate interactions, a distinctive peak of the square of the autocorrelation function is visible.
	Black curves are computed from the asymptotic formula \eqref{delt}.
	(b) Entanglement entropy averaged over the interval $t \in [0,T_{\text{rev}}]$,  $\langle \mathcal{E} \rangle _{T_{\text{rev}}}$ for $N=2$ bosons.
	Right inset provides a short time behavior of the entanglement entropy.
	Interactions induce a growth of entropy in time.
    For strong enough interactions entropy saturate at some value that does not further change.
	In the fermionized regime, average value of entropy does not change in time with minimal values reached for multiples of $T_{\text{rev}}/2$. Left inset shows evolution of entropy divided by its initial value for different interaction strengths. We observe the rapid growth of that quantity for intermediate interactions for which time coherence is lost over time.
	One can see that interaction introduces temporal incoherence in the system, which is later suppressed while entering coherent, fermionized regime.
	}
\end{figure}

\begin{figure}[t]
	\includegraphics[width=1.\linewidth]{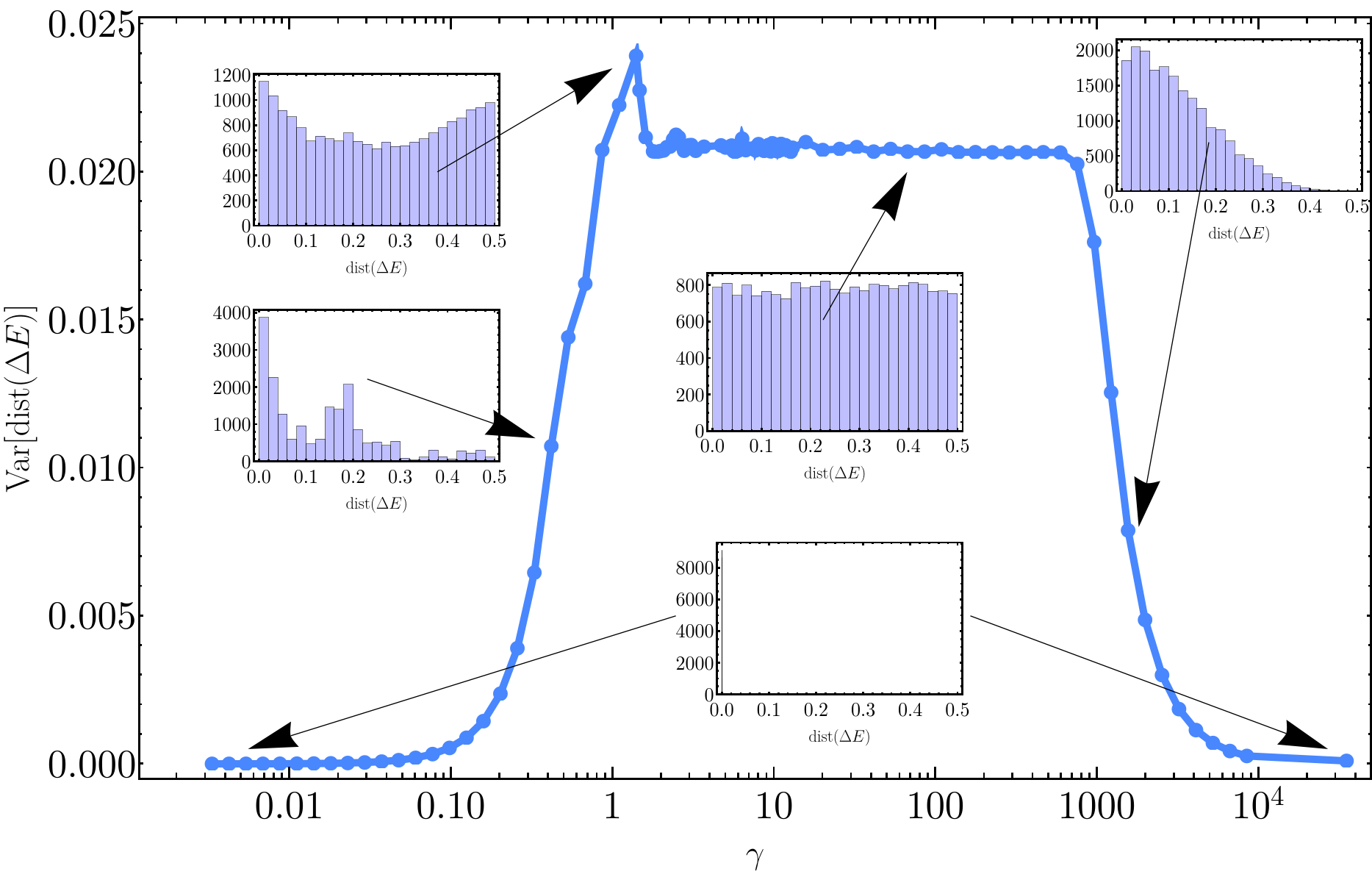}
	\caption{\label{fig5} Variance of $\text{dist}(\Delta E)$ (see the main text) for $N=3$ bosons as a function of the interaction with histograms illustrating different regimes.
	For weak and strong interactions, the histograms are very narrow and the corresponding variance is almost zero, as the energies of the system are close to being commensurate.
	The nonzero values of variance correspond to energy differences deviating from the form $\text{(integer)} \times 2 \pi \frac{\hbar}{T_{\text{rev}}}$.
	Such a property of the spectrum affects temporal coherence of the system by destroying perfect revivals and shifting the revival time.
	Around $\gamma \sim 1.5$ we observe a small peak followed by a plateau at the level given by the variance of uniform distribution. 
	}
\end{figure}

\section{Dephasing signatures}
We now proceed to introduce measures that allow us to quantify the coherence and its loss due to the finite interaction in the system.
First, we will consider a square of an autocorrelation function~\cite{Robinett2004}:
\begin{align} \label{autocorr}
A(t)=|\langle \Psi(t)| \Psi (t=0)\rangle|^2,
\end{align}
that quantifies the coherence over time, manifesting how close the initial wave function is reproduced near the revival time.
Then, we will analyze the time at which the autocorrelation function achieves maximum close to the revival time, $T_\text{rev}^{\gamma} = T_\text{rev} (1+\delta \tau)$.


First, we analyze the behavior of a revival time as a function of interaction strength (see Fig.~\ref{fig4}(a)). 
By plotting $A(t)$ close to the revival time $T_{\text{rev}}$ (insets in Fig.~\ref{fig4}(a)) for different interaction strengths, one can see that for a weak interaction perfect revival is destroyed as the maximum value of $A$ decreases with the interaction strength.
Nonetheless, the revival time is not shifted and remains at $T_{\text{rev}}$.
At the fermionization, system is again perfectly revived at $T_{\text{rev}}$, however when the interaction strength becomes smaller, imperfect revival occurs at times slightly larger than $T_{\text{rev}}$.

In the Appendix~\ref{appc}, we evaluate this behavior analytically by a perturbative expansion for the weakly interacting and nearly fermionized regimes.
In the latter case, we find a universal result 
\begin{equation}
\label{delt}
    \delta \tau = \frac{4(N-1)}{N\gamma} +\text{O}\left(\left(1/\gamma\right)^2\right),
\end{equation}
which gives a correction to the revival time for an arbitrary state in the box, as the result does not depend on the overlaps between the initial state and the eigenstates of the whole box.
As for the weakly interacting gas, we have shown that the linear correction vanishes for an arbitrary initial state of two atoms.
As Bethe equations in this regime cannot be analytically solved~\cite{Syrwid2020}, the explicit expansion for higher number of atoms was not done.
Nonetheless, results for 3 atoms suggest the same behavior for larger numbers of atoms.

Additionally, for intermediate interactions, another imperfect revival occurs, however its maximal value is smaller for three atoms than for two of them, suggesting that it may disappear in the limit of large number of atoms.

Secondly, we look at the entanglement entropy (see Fig.~\ref{fig4}(b)):
\begin{align}
\mathcal{E} (t) = - \sum_j \lambda_j (t) \log \lambda_j (t),
\end{align}
where $\lambda_j$ are eigenvalues of a single-particle density matrix. 
This observable quantifies the number of natural orbitals with a significant contribution during the evolution.
For a noninteracting system it remains constant during the time evolution.
In a weak-to-moderate interaction regime, the entanglement entropy rises rapidly in comparison to its initial value (left inset in Fig.~\ref{fig4}(b)) and for strong enough interaction, it saturates at some value at times larger than the revival time (right inset in Fig.~\ref{fig4}(b)).
The situation is different for weak interactions, where we have checked that entropy does not saturate even at very long times.
On the other hand, for a very strong interaction, system gets closer to the fermionization regime in which the entanglement entropy exhibit stable, oscillatory behavior, characteristic for the Tonks-Girardeau wave function.
From these observations we conclude that the growth of entropy in time corresponds to disappearence of revivals visible in the square of autocorrelation function.
Its value, averaged over one revival time, is not a monotonic function of the interaction strength---it has a distinctive maximum at some intermediate interaction, $\gamma \approx 15$.

Thirdly, we point out the relevance of the distribution of eigenenergies of the system.
Both in the ideal and fermionized cases, energies may be written as integer multiples of the ground state energy of box potential $\frac{\hbar^2 \pi^2}{2mL^2}=2 \pi \frac{\hbar}{T_{\text{rev}}}$.
Such a structure of the spectrum guarantees the revival of the initial state after multiples of revival time $T_{\text{rev}}$ as all phase factors visible in the expression are then equal to unity.
This is not the case for the system with finite interactions, where energies shift away from that form giving rise to dephasing and imperfect revivals.
Now, it is clear from expressions for expectation values of observables during their time evolution that one should look for aforementioned structure in the set of energy differences rather than in the bare energies of the system.
To quantify deviation of energy differences from the form $(\text{integer}) \times 2 \pi \frac{\hbar}{T_{\text{rev}}}$, we proceed as follows.
For each energy difference $\Delta E$, we consider the distance $\text{dist} (\Delta E)$ to the closest energy of the form $(\text{integer}) \times 2 \pi \frac{\hbar}{T_{\text{rev}}}$. 
Then we collect the data for energy differences stemming from 200 lowest energies in the spectrum and make a histogram.
We present these results in the Fig.~\ref{fig5}.
As expected, both in the ideal and fermionized regimes, histograms take the form of very narrow peak located near zero.
For finite interactions, the histograms become broader, contributing to the loss of the temporal coherence in the system.
To further study that problem across a wide range of interaction values, we plot variance of the data as a function of $\gamma$.
As expected, in weakly interacting and fermionized regimes, the variance is very small, contrary to the intermediate regime that corresponds to loss of the temporal coherence in the system.

Each of these signatures---regularity of a quantum carpet pattern both in spatial and momentum representation, presence of the quantum revival, incommensurateness of energy differences and nonmonotonic behavior of the time-averaged entanglement entropy---heralds the loss of temporal coherence in the system due to interaction and then its reappearance in the strongly interacting regime.
In previous studies, it has been shown that a quantum carpet design is a purely interference effect and as such it can serve as an alternative probe for decoherence effects that does not rely on the reconstruction of the Wigner function~\cite{Kazemi2013}.
In such an approach, the decoherence present in the system is probed via access to the single-particle density, which is well suited to e.g. experiments involving diffracton gratings that can reproduce similar revival dynamics.

Here, we present a model in which temporal coherence of the system is affected by the interparticle interaction and a quantum carpet pattern is accordingly smeared out.
The particular excitation scheme we consider can be achieved in current experimental settings, with box traps~\cite{Gaunt2013} and fast geometric quench techniques~\cite{Wilson2020}, also in quasi-one-dimensional geometries with tunable short range interactions.
Moreover, coherent diffraction has been achieved also in systems other than ultracold gases, involving atoms~\cite{Kurtsiefer1997}, large molecules~\cite{Hornberger2003,Hornberger2012a}, light~\cite{Chapman1995a} and electrons~\cite{Sonnentag2007a}.
Quantum carpets may provide a natural tool to investigate different decoherence scenarios in these systems, among others due to interaction as presented in here.
These other decoherence processes may include e.g. particle losses or imperfections of the experimental setups.

\section{Conclusions and outlook}
We have analyzed exactly a single- to many-body crossover of a highly nonequilibrium state by investigating a spatiotemporal design known as a quantum carpet.
The latter can be realized in plethora of systems, both for matter and light waves and might be utilized as a decoherence probe that provides a direct access to the Wigner function of the considered state.
We have shown that interparticle interaction not only destroys the coherence in the system, but, if strong enough, can make the system coherent again.
Such a behavior is possible due to a particular excitation scheme that allows a quantum carpet of fermionized bosons to preserve regularity.
We have provided a thorough analysis of the crossover between bosonic and fermionic quantum carpets that emerge under such a scenario.

As a future line of work, quantum carpets can be investigated in other many-body scenarios involving different interactions encountered in atomic systems, such as attractive or dipolar potentials.
Moreover, two- and three-dimensional geometries would also produce different type of decoherence and affect the quantum carpet design differently.

\begin{acknowledgments}
M. \L{}. acknowledges the support from the (Polish) National Science Center Grant 2018/31/N/ST2/01429. 
P. T. G. is financed from the (Polish) National Science Center Grants 2018/31/N/ST2/01429 and 2020/36/T/ST2/00065. 
K. Rz. is supported from the (Polish) National Science Center Grant 2018/29/B/ST2/01308.
Center for Theoretical Physics of the Polish Academy of Sciences is a member of the National Laboratory of Atomic, Molecular and Optical Physics (KL FAMO).
\end{acknowledgments}

\appendix
\section{Deriving $\rho(x,x',t)$}\label{appa}
We now proceed to evaluate reduced single-particle density matrix for noninteracting fermions.
If $\phi _{i}\left( x',t\right)$ is a time-dependent natural orbital (in ideal fermionic case natural orbitals are just initial single particle orbitals), then the reduced single-particle density matrix reads
\begin{equation}
\rho \left( x,x',t\right) =\sum ^{\infty }_{i=0}\mu _{i} \ \phi_i^{\ast }\left(x,t\right) \phi _{i}\left( x',t\right), \end{equation}
where $\mu_i = 1/N$ are equal weights.
We can define parts $\rho_i$ of the full density matrix, associated with a given initial orbital:
\begin{align}
\rho\left( x,x',t\right) =\sum _{i=0}\mu _{i} \ \rho_{\left( i\right)} \left( x,x',t\right), \nonumber \\ 
\rho_{\left( i\right)} \left( x,x',t\right)  =\phi_i^{\ast }\left( x,t\right) \phi_{i}\left( x',t\right) 
\end{align}
After the release from the initial confinement, it can be then expanded in single-particle orbitals of the larger box:
\begin{align}
\rho_{\left( i\right)} \left( x,x',t\right) &=\left( \sum_{k}\Lambda_{i,k}^{\ast }\varphi_{k}^{\ast }\left( x\right) e^{iE_{k}t/\hbar }\right) \\ \nonumber 
& \times \left( \sum _{l}\Lambda_{i,l}\varphi_{l}\left( x'\right) e^{-iE_{l}t/\hbar }\right) \\ \nonumber
&=\sum _{k,l}\Lambda_{i,k}^{\ast }\Lambda_{i,l} \varphi_k \left( x\right) \varphi_{l}\left( x'\right)  e^{i\left( E_k-E_l\right) t/\hbar }.
\end{align}
The real part of a given contribution can be written as
\begin{align}
& Re \ \rho_{(i)}= \sum _{k,l} \Lambda_{i,k} \Lambda_{i,l}\varphi_{k}\left( x\right) \varphi_{l}\left( x'\right) \ \cos \left( \frac{t}{\hbar} \left( E_{k}-E_{l}\right) \right) \\ \nonumber
&=\sum_{k}\left| \Lambda_{ik}\right|^{2} \varphi_{k}\left( x\right) \varphi_{k}\left( x'\right) \\ \nonumber
&+\sum_{k=1,l=1} \Lambda _{i,k} \Lambda _{i,k+l} \varphi _{k}\left( x\right) \varphi _{k+l}\left( x'\right) \ \cos \left( \frac{t}{\hbar}  \left( E_{k}-E_{k+l}\right) \right) \\ \nonumber
&+\sum_{k=1,l=1} \Lambda _{i,k} \Lambda _{i,k+l} \varphi _{k}\left( x'\right) \varphi _{k+l}\left( x\right) \ \cos \left( \frac{t}{\hbar}  \left( E_{k}-E_{k+l}\right) \right) \\ \nonumber
&=\sum _{k} I_{kk}+\sum_{k=1,l=1} \frac{1}{2L}\Lambda _{i,k}\Lambda _{i,k+l}\left( I_{kl}\left( x,x'\right) +I_{kl}\left( x',x\right) \right), 
\end{align}
where
\begin{equation}
I_{kl}\left( x,x'\right) =2 L \varphi _{k} \left( x\right) \varphi_{k+l}\left( x'\right)  \cos \left( l\left( l+2k\right)  \frac{\pi }{L} v_{0} t\right).
\end{equation}
Analogously for the imaginary part:
\begin{align}
Im \ \rho_{(i)}
=\sum_{k=1,l=1} \frac{1}{2L}\Lambda _{i,k}\Lambda _{i,k+l}\left( J_{kl}\left( x,x'\right) -J_{kl}\left( x',x\right) \right), 
\end{align}
with
\begin{equation}
J_{kl}\left( x,x'\right) = - 2 L \varphi _{k} \left( x\right) \varphi_{k+l}\left( x'\right)  \sin \left( l\left( l+2k\right)  \frac{\pi }{L} v_{0} t\right) .
\end{equation}
The function $I_{kl}$ can be evaluated as
\begin{widetext}
\begin{align}
I_{kl}&=4 \sin \left( \frac{k\pi x}{L}\right)  \sin \left( \frac{\left( k+1\right)  \pi  x'}{L}\right)  \cos \left( l\left( l+2k\right)  \frac{\pi }{L}v_{0}t\right) \\ \nonumber
&= \cos\left( \frac{\pi }{L} l u\right) \left(  \cos \frac{\pi }{L} \left(  l+2k\right) \left( v - l v_{0}t \right) + \cos \frac{\pi }{L} \left(  l+2k\right) \left( v + l v_{0}t \right)  \right) \\ \nonumber
&- \cos\left( \frac{\pi }{L} l v\right) \left(  \cos \frac{\pi }{L} \left(  l+2k\right) \left( u - l v_{0}t \right) + \cos \frac{\pi }{L} \left(  l+2k\right) \left( u + l v_{0}t \right)  \right) \\ \nonumber
&- \sin\left( \frac{\pi }{L} l u\right) \left(  \sin \frac{\pi }{L} \left(  l+2k\right) \left( v - l v_{0}t \right) + \sin \frac{\pi }{L} \left(  l+2k\right) \left( v + l v_{0}t \right)  \right) \\ \nonumber
&+ \sin\left( \frac{\pi }{L} l v\right) \left(  \sin \frac{\pi }{L} \left(  l+2k\right) \left( u - l v_{0}t \right) + \sin \frac{\pi }{L} \left(  l+2k\right) \left( u + l v_{0}t \right)  \right) 
\end{align}
\end{widetext}
where
\begin{equation}
v=\frac{x+x'}{2}, \ \ \ u=\frac{x-x'}{2}, 
\end{equation}
With the following identities
\begin{align}
\tilde{I_{kl}}\left( u,v\right) =I_{kl}\left( u-v,u+v\right)  = I_{kl} \left( x,x'\right), \nonumber \\
I_{kl}\left( x, x'\right) + I_{kl}\left( x',x\right)  =  \tilde{I_{kl}}\left( u,v\right) +\tilde{I_{kl}} \left( u,-v\right),
\end{align}
we can immediately write
\begin{widetext}
\begin{align}
&\tilde{I_{kl}}\left( u,v\right) +\tilde{I_{kl}} \left( u,-v\right) = \\ \nonumber
& 2\cos\left( \frac{\pi }{L} lu\right) \left(  \cos \frac{\pi }{L} \left(  l+2k\right) \left( v - l v_{0}t \right) + \cos \frac{\pi }{L} \left(  l+2k\right) \left( v + l v_{0}t \right)  \right) \\ \nonumber
- &2\cos\left( \frac{\pi }{L} lv\right) \left(  \cos \frac{\pi }{L} \left(  l+2k\right) \left( u - l v_{0}t \right) + \cos \frac{\pi }{L} \left(  l+2k\right) \left( u + l v_{0}t \right)  \right). 
\end{align}
Then, we can reorganize terms to get
\begin{align}
Re \ \rho_{\left( i\right)} =&-\sum _{p\in \mathbb{Z} }\sum ^{\infty }_{k=1} \frac{1}{L}\Lambda_{i,k}\Lambda_{i,k+ \left| p\right|} \cos \left( \frac{\pi }{L}  \left| p\right| v\right)  \cos \frac{\pi }{L}\left( \left| p\right| +2k\right)  \left( u-p v_{0}t\right)  \nonumber \\
&+ \sum _{p \in \mathbb{Z} }\sum^{\infty }_{k=1} \frac{1}{L}\Lambda_{i,k}\Lambda_{i,k+ \left| p\right|} \cos \left( \frac{\pi }{L} \ \left| p\right| u\right)  \cos \frac{\pi }{L}\left( \left| p\right| +2k\right)  \left( v-p v_{0}t\right) 
\end{align}
By analogous procedure for $J_{kl}$ we get
\begin{align}
Im \ \rho_{\left( i\right)} =&-\sum _{p\in \mathbb{Z} }\sum ^{\infty }_{k=1} \frac{1}{L}\Lambda_{i,k}\Lambda_ {i,k+ \left| p\right|} sgn(p)  \sin \left( \frac{\pi }{L}  \left| p\right| v\right)  \cos \frac{\pi }{L}\left( \left| p\right| +2k\right)  \left( u-p v_{0}t\right)  \nonumber \\
&+ \sum _{p\in \mathbb{Z} }\sum ^{\infty }_{k=1} \frac{1}{L}\Lambda_{i,k}\Lambda_ {i,k+ \left| p\right|} sgn(p)  \sin \left( \frac{\pi }{L} \ \left| p\right| u\right)  \cos \frac{\pi }{L}\left( \left| p\right| +2k\right)  \left( v-pv_{0}t\right).
\end{align}
It allows to write the whole reduced single-particle density matrix in a concise form:
\begin{equation}
\rho \left( x,x',t\right) =\sum ^{\infty }_{i=1}\sum _{\pm }\sum _{p\in \mathbb{Z} }\sum ^{\infty }_{k=1} \pm \mu_{i} \frac{1}{L}\Lambda_{i,k}\Lambda_ {i,k+ \left| p\right|} e^{i \frac{\pi}{L} p \frac{x' \pm x}{2}} \cos \left[\frac{\pi }{L}\left( \left| p\right| +2k\right)  \left( \frac{x' \mp x}{2}-pv_{0}t\right) \right].
\end{equation}

\section{Bethe states calculations}\label{appb}
Here we present details of calculation of normalization constants, overlaps and single-particle density matrix.
In the remaining part we will often encounter $N$-dimensional integrals and it is useful to note that for Bose-symmetric function $F(\{x\})$ integration over N dimensional hypercube $L^N$ can be transformed~\cite{Staron2020} into integration over the fundamental domain $0\leq x_1\leq \ldots\leq x_N\leq L$
\begin{equation}
   \int_0^L \dd x_N \int_0^L \dd x_{N-1} \ldots \int_0^L \dd x_1 \, F(\{x\}) = N! \int_0^L \dd x_N \int_0^{x_N} \dd x_{N-1}\ldots \int_0^{x_2} \dd x_1 \, F(\{x\}).
\end{equation}
We use that property to calculate normalization constants
\begin{equation}
\label{norm}
\begin{aligned}
   \mathcal{N}_{\{k\}} = &N!\sum_{\{\sigma\}}\sum_{\{\epsilon\}}\sum_Q \sum_P \sigma_1 \epsilon_1 \ldots \sigma_N \epsilon_N \prod_{i<j}\bigg(1+\frac{ic}{\sigma_i k_i+ \sigma_j k_j}\bigg)\bigg(1-\frac{ic}{\epsilon_i k_i+ \epsilon_j k_j}\bigg) \bigg(1-\frac{ic}{\sigma_{Qi}k_{Qi}-\sigma_{Qj}k_{Qj}} \bigg)\times\\
   &\bigg(1+\frac{ic}{\epsilon_{Pi}k_{Pi}-\epsilon_{Pj}k_{Pj}} \bigg) I_N\Big(\epsilon_{P1}k_{P1}-\sigma_{Q1}k_{Q1},\ldots,\epsilon_{PN}k_{PN}-\sigma_{QN}k_{QN},L\Big),
\end{aligned}
\end{equation}
where we have introduced
\begin{equation}
\label{intI}
    I_N\Big(\alpha_1,\ldots,\alpha_N,L\Big)=\int_0^L \dd x_N \int_0^{x_N} \dd x_{N-1}\ldots \int_0^{x_2} \dd x_1 \exp \Big[i(\alpha_1 x_1+\ldots+\alpha_N x_N)\Big].
\end{equation}
The property \eqref{bosesym} allowed us to get rid of sign functions appearing in wavefunctions. The integrals \eqref{intI} can be calculated analytically. Coefficients $C_{\{k\}}$ are calculated in the following way
\begin{equation}
\begin{aligned}
    C_{\{k\}} =&\frac{N!}{\sqrt{\mathcal{N}_{\{k^0\}} \mathcal{N}_{\{k\}}}}\sum_{\{\sigma\}}\sum_{\{\epsilon\}}\sum_Q \sum_P \sigma_1 \epsilon_1 \ldots \sigma_N \epsilon_N \prod_{i<j}\bigg(1+\frac{ic}{\sigma_i k_i+ \sigma_j k_j}\bigg)\bigg(1-\frac{ic}{\epsilon_i k^0_i+ \epsilon_j k^0_j}\bigg) \bigg(1-\frac{ic}{\sigma_{Qi}k_{Qi}-\sigma_{Qj}k_{Qj}} \bigg)\times\\
   &\bigg(1+\frac{ic}{\epsilon_{Pi}k^0_{Pi}-\epsilon_{Pj}k^0_{Pj}} \bigg) I_N\Big(\epsilon_{P1}k^0_{P1}-\sigma_{Q1}k_{Q1},\ldots,\epsilon_{PN}k_{PN}^0-\sigma_{QN}k_{QN},D\Big).
\end{aligned}
\end{equation}
Where $k^0_1, \ldots, k^0_N$ are solutions of 
\begin{equation}
    k_i^0 D = \pi n_i +\sum_{j \neq i} \Bigg( \arctan \frac{c}{k_i^0-k_j^0} +\arctan \frac{c}{k_i^0+k_j^0} \Bigg).
\end{equation}
Next, we go to single-particle density
\begin{equation}
    \rho(x,x',t)= \int_0^L \dd x_N \int_0^L \dd x_{N-1} \ldots \int_0^L \dd x_2  \Psi^*(x,\ldots,x_N,t) \Psi(x',\ldots,x_N,t).
\end{equation}
We may use property \eqref{bosesym} to simplify the integral
\begin{equation}
    \rho(x,x',t)= (N-1)!\int_0^L \dd x_N \int_0^{x_N} \dd x_{N-1}\ldots \int_0^{x_3} \dd x_2 \Psi^*(x,\ldots,x_N,t) \Psi(x',\ldots,x_N,t).
\end{equation}

Explicit form of time-evolved single-particle density reads
\begin{equation}
\label{densitymatrix}
    \begin{aligned}
        &\rho(x,x',t)=(N-1)!\sum_u \sum_v \frac{1}{\sqrt{\mathcal{N}_u \mathcal{N}_v}}\, e^{i(E_v-E_u)t}\, C_v^*C_u\sum_{\{\sigma\}}\sum_{\{\epsilon\}}\sum_Q \sum_P\sigma_1 \epsilon_1 \ldots \sigma_N \epsilon_N \prod_{l>1} \bigg(1+\frac{ic}{\sigma_1 k^v_1+ \sigma_l k^v_l}\bigg) \times\\
        &\bigg(1-\frac{ic}{\epsilon_1 k^u_1+ \epsilon_l k^u_l}\bigg) \prod_{1<i<j}\bigg(1+\frac{ic}{\sigma_i k^v_i+ \sigma_j k^v_j}\bigg) \bigg(1-\frac{ic}{\epsilon_i k^u_i+ \epsilon_j k^u_j}\bigg) \bigg(1-\frac{ic}{\sigma_{Qi}k^v_{Qi}-\sigma_{Qj}k^v_{Qj}} \bigg)\bigg(1+\frac{ic}{\epsilon_{Pi}k^u_{Pi}-\epsilon_{Pj}k^u_{Pj}} \bigg) \times \\
        & \Bigg( \int_0^L \dd x_N \int_0^{x_N} \dd x_{N-1}\ldots \int_0^{x_3} \dd x_2 \prod_{m>1} \bigg(1-\frac{ic \, \text{sgn}(x_m-x)}{\sigma_{Q1}k^v_{Q1}-\sigma_{Qm}k^v_{Qm}} \bigg) \bigg(1+\frac{ic\, \text{sgn}(x_m-x')}{\epsilon_{P1}k^u_{P1}-\epsilon_{Pm}k^u_{Pm}} \bigg) \times \\
        & \exp\Big[i[(\epsilon_{P2}k^u_{P2}-\sigma_{Q2}k^v_{Q2})x_2+\ldots+(\epsilon_{PN}k^u_{PN}-\sigma_{QN}k^v_{QN})x_N]\Big]\Bigg) \exp[i(\epsilon_{P1}k^u_{P1}x'-\sigma_{Q1}k^v_{Q1}x)].
    \end{aligned}
\end{equation}
The integrals involving sign functions can be calculated analytically upon dividing the integration region into parts where sign functions do not change its values. Thus, all results for single-particle density in the position space presented in our paper are obtained from evaluation of analytical expressions. Having full matrix \eqref{densitymatrix} calculated for a finite number of points $N_{grid}=101$ we diagonalize it obtaining eigenvalues $\lambda_j(t)$ used to calculate entropy
\begin{equation}
    \mathcal{E}(t) = -\sum_j \lambda_j(t) \log \lambda_j(t).
\end{equation}
\end{widetext}

\section{Perturbative evaluation of the revival time}\label{appc}
We aim to expand autocorrelation function perturbatively for nearly fermionized and weakly interacting gas.
In the limiting cases of fermionized and noninteracting gas, respectively, $N$-body eigenstate $\ket{n}$ can unambiguously be identified by a set of positive numbers $\{k_{ni}\}$, $i\in \{1, \dots, N\}$ and has energy $E_n = \frac{\hbar^2 \pi^2}{2 m L^2}\sum_i k_{ni}^2$.
For simplicity, let us introduce a number $M_n$:
\begin{equation}
T_{\text{rev}} E_n / \hbar = 2 \pi \sum_{i} k_{ni}^2 = 2 \pi M_n,
\end{equation}
where the noninteracting revival time reads $T_{\text{rev}}= 4 L^2 m/\pi \hbar$.
If there is finite interaction introduced, the energies start to slightly deviate, with a linear correction $\delta E_n$ reading
\begin{equation}
T_{\text{rev}} \delta E_n / \hbar = 4 \pi \sum_{i} k_{ni} \delta k_{ni} = 4 \pi \delta M_n,
\end{equation}
where we introduced another number $\delta M_n$.
We will now expand the autocorrelation function~\eqref{autocorr} close to the $T_{\text{rev}}$:
\begin{widetext}
\begin{align}
    &|A(T_{\text{rev}}(1+\delta \tau))|^2 \approx \sum_n |a_n|^4 + 2 \sum_{m\neq n} |a_n|^2|a_m|^2 \nonumber \\
    &-2 \sum_{m\neq n} |a_n|^2|a_m|^2 \left( 2 \pi^2 \left(M_n - M_m\right)^2 \delta \tau^2 + 8 \pi^2 \left(M_n - M_m\right) \left(\delta M_n - \delta M_m\right) \delta \tau - 16 \pi^2 \delta M_n \delta M_n \right),
\end{align}
\end{widetext}
where we expanded up to second order in $\delta \tau$ and up to first order in $\delta M_n$ and $a_n = \bra{\Psi_0}\ket{n}$ are the overlaps of the initial state with the eigenstates of the Hamiltonian.
We need to maximize it in order to find a time at which imperfect revival happens.
As it is a quadratic function of $\delta \tau$, it can be readily done, yielding
\begin{widetext}
\begin{align}
\delta \tau = -\frac{2 \sum_{m\neq n} |a_n|^2|a_m|^2 \left(M_n - M_m\right) \left(\delta M_n - \delta M_m\right)  }{\sum_{m\neq n} |a_n|^2|a_m|^2 \left(M_n - M_m\right)^2}.
\end{align}
\end{widetext}
In case of the nearly fermionized gas, one can find analytical formula~\cite{Syrwid2020}:
\begin{align}\label{expk}
\delta k_{ni} = - k_{ni} \frac{2(N-1)}{N \gamma},
\end{align}
from which follows that
\begin{align}
\delta M_n - \delta M_m = -\frac{2(N-1)}{N \gamma} \left(M_n - M_m \right),
\end{align}
which gives the time of the imperfect revival
\begin{equation}
    \delta \tau = \frac{4(N-1)}{N\gamma} +\text{O}\left(\left(1/\gamma\right)^2\right).
\end{equation}
Note that this result does not depend on the overlaps $a_n$ and as such, it is universal for every initial state.
In case of a weakly interacting gas, there is no analytical equivalent of formula~\eqref{expk}, however Bethe equations can be expanded in a case of a two-body problem, giving
\begin{align}
\delta k_{n1} &= \frac{k_{n1}}{ k_{n1}^2 - k_{n2}^2} 2 N \gamma \nonumber \\
\delta k_{n2} &= -\frac{k_{n2}}{ k_{n1}^2 - k_{n2}^2} 2 N \gamma 
\end{align}
for $k_{n1} > k_{n2}$.
Analogous relation can be computed for $k_{n1} < k_{n2}$, altogether yielding $\delta M_n - \delta M_m =\text{O}\left(\gamma^2\right) $ and finally
\begin{align}
 \delta \tau =\text{O}\left(\gamma^2\right).
\end{align}
Note that our numerical results suggest that similar formula holds also for larger numbers of atoms.

\bibliography{FQC2,Crossover}

\begin{thebibliography}{63}%
\makeatletter
\providecommand \@ifxundefined [1]{%
 \@ifx{#1\undefined}
}%
\providecommand \@ifnum [1]{%
 \ifnum #1\expandafter \@firstoftwo
 \else \expandafter \@secondoftwo
 \fi
}%
\providecommand \@ifx [1]{%
 \ifx #1\expandafter \@firstoftwo
 \else \expandafter \@secondoftwo
 \fi
}%
\providecommand \natexlab [1]{#1}%
\providecommand \enquote  [1]{``#1''}%
\providecommand \bibnamefont  [1]{#1}%
\providecommand \bibfnamefont [1]{#1}%
\providecommand \citenamefont [1]{#1}%
\providecommand \href@noop [0]{\@secondoftwo}%
\providecommand \href [0]{\begingroup \@sanitize@url \@href}%
\providecommand \@href[1]{\@@startlink{#1}\@@href}%
\providecommand \@@href[1]{\endgroup#1\@@endlink}%
\providecommand \@sanitize@url [0]{\catcode `\\12\catcode `\$12\catcode
  `\&12\catcode `\#12\catcode `\^12\catcode `\_12\catcode `\%12\relax}%
\providecommand \@@startlink[1]{}%
\providecommand \@@endlink[0]{}%
\providecommand \url  [0]{\begingroup\@sanitize@url \@url }%
\providecommand \@url [1]{\endgroup\@href {#1}{\urlprefix }}%
\providecommand \urlprefix  [0]{URL }%
\providecommand \Eprint [0]{\href }%
\providecommand \doibase [0]{https://doi.org/}%
\providecommand \selectlanguage [0]{\@gobble}%
\providecommand \bibinfo  [0]{\@secondoftwo}%
\providecommand \bibfield  [0]{\@secondoftwo}%
\providecommand \translation [1]{[#1]}%
\providecommand \BibitemOpen [0]{}%
\providecommand \bibitemStop [0]{}%
\providecommand \bibitemNoStop [0]{.\EOS\space}%
\providecommand \EOS [0]{\spacefactor3000\relax}%
\providecommand \BibitemShut  [1]{\csname bibitem#1\endcsname}%
\let\auto@bib@innerbib\@empty
\bibitem [{\citenamefont {Talbot}(1836)}]{Talbot1836}%
  \BibitemOpen
  \bibfield  {author} {\bibinfo {author} {\bibfnamefont {H.~F.}\ \bibnamefont
  {Talbot}},\ }\bibfield  {title} {\bibinfo {title} {{Facts relating to optical
  science}},\ }\href@noop {} {\bibfield  {journal} {\bibinfo  {journal}
  {Philos. Mag.}\ }\textbf {\bibinfo {volume} {9}},\ \bibinfo {pages} {401}
  (\bibinfo {year} {1836})}\BibitemShut {NoStop}%
\bibitem [{\citenamefont {{Lord Rayleigh}}(1881)}]{LordRayleigh1881}%
  \BibitemOpen
  \bibfield  {author} {\bibinfo {author} {\bibnamefont {{Lord Rayleigh}}},\
  }\bibfield  {title} {\bibinfo {title} {{On copying diffraction gratings and
  on some phenomenon connected therewith}},\ }\href@noop {} {\bibfield
  {journal} {\bibinfo  {journal} {Philos. Mag.}\ }\textbf {\bibinfo {volume}
  {11}},\ \bibinfo {pages} {196} (\bibinfo {year} {1881})}\BibitemShut
  {NoStop}%
\bibitem [{\citenamefont {Berry}(1996)}]{Berry1996}%
  \BibitemOpen
  \bibfield  {author} {\bibinfo {author} {\bibfnamefont {M.~V.}\ \bibnamefont
  {Berry}},\ }\bibfield  {title} {\bibinfo {title} {{Quantum fractals in
  boxes}},\ }\href {https://doi.org/10.1088/0305-4470/29/20/016} {\bibfield
  {journal} {\bibinfo  {journal} {J. Phys. A. Math. Gen.}\ }\textbf {\bibinfo
  {volume} {29}},\ \bibinfo {pages} {6617} (\bibinfo {year}
  {1996})}\BibitemShut {NoStop}%
\bibitem [{\citenamefont {W{\'{o}}jcik}\ \emph {et~al.}(2000)\citenamefont
  {W{\'{o}}jcik}, \citenamefont {Bia{\l}ynicki-Birula},\ and\ \citenamefont
  {{\.{Z}}yczkowski}}]{Wojcik2000}%
  \BibitemOpen
  \bibfield  {author} {\bibinfo {author} {\bibfnamefont {D.}~\bibnamefont
  {W{\'{o}}jcik}}, \bibinfo {author} {\bibfnamefont {I.}~\bibnamefont
  {Bia{\l}ynicki-Birula}},\ and\ \bibinfo {author} {\bibfnamefont
  {K.}~\bibnamefont {{\.{Z}}yczkowski}},\ }\bibfield  {title} {\bibinfo {title}
  {{Time Evolution of Quantum Fractals}},\ }\href
  {https://doi.org/10.1103/PhysRevLett.85.5022} {\bibfield  {journal} {\bibinfo
   {journal} {Phys. Rev. Lett.}\ }\textbf {\bibinfo {volume} {85}},\ \bibinfo
  {pages} {5022} (\bibinfo {year} {2000})}\BibitemShut {NoStop}%
\bibitem [{\citenamefont {Gao}\ \emph {et~al.}(2019)\citenamefont {Gao},
  \citenamefont {Zhai},\ and\ \citenamefont {Shi}}]{Gao2019}%
  \BibitemOpen
  \bibfield  {author} {\bibinfo {author} {\bibfnamefont {C.}~\bibnamefont
  {Gao}}, \bibinfo {author} {\bibfnamefont {H.}~\bibnamefont {Zhai}},\ and\
  \bibinfo {author} {\bibfnamefont {Z.-Y.}\ \bibnamefont {Shi}},\ }\bibfield
  {title} {\bibinfo {title} {Dynamical fractal in quantum gases with discrete
  scaling symmetry},\ }\href {https://doi.org/10.1103/PhysRevLett.122.230402}
  {\bibfield  {journal} {\bibinfo  {journal} {Phys. Rev. Lett.}\ }\textbf
  {\bibinfo {volume} {122}},\ \bibinfo {pages} {230402} (\bibinfo {year}
  {2019})}\BibitemShut {NoStop}%
\bibitem [{\citenamefont {Buchkremer}\ \emph {et~al.}(2000)\citenamefont
  {Buchkremer}, \citenamefont {Dumke}, \citenamefont {Levsen}, \citenamefont
  {Birkl},\ and\ \citenamefont {Ertmer}}]{Buchkremer2000}%
  \BibitemOpen
  \bibfield  {author} {\bibinfo {author} {\bibfnamefont {F.~B.~J.}\
  \bibnamefont {Buchkremer}}, \bibinfo {author} {\bibfnamefont
  {R.}~\bibnamefont {Dumke}}, \bibinfo {author} {\bibfnamefont
  {H.}~\bibnamefont {Levsen}}, \bibinfo {author} {\bibfnamefont
  {G.}~\bibnamefont {Birkl}},\ and\ \bibinfo {author} {\bibfnamefont
  {W.}~\bibnamefont {Ertmer}},\ }\bibfield  {title} {\bibinfo {title} {{Wave
  Packet Echoes in the Motion of Trapped Atoms}},\ }\href
  {https://doi.org/10.1103/PhysRevLett.85.3121} {\bibfield  {journal} {\bibinfo
   {journal} {Phys. Rev. Lett.}\ }\textbf {\bibinfo {volume} {85}},\ \bibinfo
  {pages} {3121} (\bibinfo {year} {2000})}\BibitemShut {NoStop}%
\bibitem [{\citenamefont {Sanz}\ and\ \citenamefont
  {Miret-Art{\'{e}}s}(2007)}]{Sanz2007}%
  \BibitemOpen
  \bibfield  {author} {\bibinfo {author} {\bibfnamefont {A.~S.}\ \bibnamefont
  {Sanz}}\ and\ \bibinfo {author} {\bibfnamefont {S.}~\bibnamefont
  {Miret-Art{\'{e}}s}},\ }\bibfield  {title} {\bibinfo {title} {{A causal look
  into the quantum Talbot effect}},\ }\href {https://doi.org/10.1063/1.2741555}
  {\bibfield  {journal} {\bibinfo  {journal} {J. Chem. Phys.}\ }\textbf
  {\bibinfo {volume} {126}},\ \bibinfo {pages} {234106} (\bibinfo {year}
  {2007})}\BibitemShut {NoStop}%
\bibitem [{\citenamefont {Heller}(1984)}]{Heller1984}%
  \BibitemOpen
  \bibfield  {author} {\bibinfo {author} {\bibfnamefont {E.~J.}\ \bibnamefont
  {Heller}},\ }\bibfield  {title} {\bibinfo {title} {{Bound-State
  Eigenfunctions of Classically Chaotic Hamiltonian Systems: Scars of Periodic
  Orbits}},\ }\href {https://doi.org/10.1103/PhysRevLett.53.1515} {\bibfield
  {journal} {\bibinfo  {journal} {Phys. Rev. Lett.}\ }\textbf {\bibinfo
  {volume} {53}},\ \bibinfo {pages} {1515} (\bibinfo {year}
  {1984})}\BibitemShut {NoStop}%
\bibitem [{\citenamefont {Kaplan}\ and\ \citenamefont
  {Heller}(1998)}]{Kaplan1998}%
  \BibitemOpen
  \bibfield  {author} {\bibinfo {author} {\bibfnamefont {L.}~\bibnamefont
  {Kaplan}}\ and\ \bibinfo {author} {\bibfnamefont {E.}~\bibnamefont
  {Heller}},\ }\bibfield  {title} {\bibinfo {title} {{Linear and Nonlinear
  Theory of Eigenfunction Scars}},\ }\href
  {https://doi.org/10.1006/APHY.1997.5773} {\bibfield  {journal} {\bibinfo
  {journal} {Ann. Phys. (N. Y).}\ }\textbf {\bibinfo {volume} {264}},\ \bibinfo
  {pages} {171} (\bibinfo {year} {1998})}\BibitemShut {NoStop}%
\bibitem [{\citenamefont {Eberly}\ \emph {et~al.}(1980)\citenamefont {Eberly},
  \citenamefont {Narozhny},\ and\ \citenamefont
  {Sanchez-Mondragon}}]{Eberly1980}%
  \BibitemOpen
  \bibfield  {author} {\bibinfo {author} {\bibfnamefont {J.~H.}\ \bibnamefont
  {Eberly}}, \bibinfo {author} {\bibfnamefont {N.~B.}\ \bibnamefont
  {Narozhny}},\ and\ \bibinfo {author} {\bibfnamefont {J.~J.}\ \bibnamefont
  {Sanchez-Mondragon}},\ }\bibfield  {title} {\bibinfo {title} {{Periodic
  Spontaneous Collapse and Revival in a Simple Quantum Model}},\ }\href
  {https://doi.org/10.1103/PhysRevLett.44.1323} {\bibfield  {journal} {\bibinfo
   {journal} {Phys. Rev. Lett.}\ }\textbf {\bibinfo {volume} {44}},\ \bibinfo
  {pages} {1323} (\bibinfo {year} {1980})}\BibitemShut {NoStop}%
\bibitem [{\citenamefont {Robinett}(2004)}]{Robinett2004}%
  \BibitemOpen
  \bibfield  {author} {\bibinfo {author} {\bibfnamefont {R.~W.}\ \bibnamefont
  {Robinett}},\ }\bibfield  {title} {\bibinfo {title} {{Quantum wave packet
  revivals}},\ }\href {https://doi.org/10.1016/j.physrep.2003.11.002}
  {\bibfield  {journal} {\bibinfo  {journal} {Phys. Rep.}\ }\textbf {\bibinfo
  {volume} {392}},\ \bibinfo {pages} {1} (\bibinfo {year} {2004})}\BibitemShut
  {NoStop}%
\bibitem [{\citenamefont {Kinzel}(1995)}]{Kinzel1995}%
  \BibitemOpen
  \bibfield  {author} {\bibinfo {author} {\bibfnamefont {W.}~\bibnamefont
  {Kinzel}},\ }\bibfield  {title} {\bibinfo {title} {{Bilder elementarer
  Quantenmechanik}},\ }\href
  {https://onlinelibrary.wiley.com/doi/pdf/10.1002/phbl.19950511215} {\bibfield
   {journal} {\bibinfo  {journal} {Phys. Bl.}\ }\textbf {\bibinfo {volume}
  {51}},\ \bibinfo {pages} {1190} (\bibinfo {year} {1995})}\BibitemShut
  {NoStop}%
\bibitem [{\citenamefont {Stifter}\ \emph {et~al.}(1997)\citenamefont
  {Stifter}, \citenamefont {Leichtie}, \citenamefont {Schleich},\ and\
  \citenamefont {Marklof}}]{Stifter1997}%
  \BibitemOpen
  \bibfield  {author} {\bibinfo {author} {\bibfnamefont {P.}~\bibnamefont
  {Stifter}}, \bibinfo {author} {\bibfnamefont {C.}~\bibnamefont {Leichtie}},
  \bibinfo {author} {\bibfnamefont {W.~P.}\ \bibnamefont {Schleich}},\ and\
  \bibinfo {author} {\bibfnamefont {J.}~\bibnamefont {Marklof}},\ }\bibfield
  {title} {\bibinfo {title} {{Das Teilchen im Kasten: Strukturen in der
  Wahrscheinlichkeitsdichte / The Particle in a Box : Structures in the
  Probability}},\ }\href {https://doi.org/10.1515/zna-1997-0501} {\bibfield
  {journal} {\bibinfo  {journal} {Zeitschrift f{\"{u}}r Naturforsch. A}\
  }\textbf {\bibinfo {volume} {52}},\ \bibinfo {pages} {377} (\bibinfo {year}
  {1997})}\BibitemShut {NoStop}%
\bibitem [{\citenamefont {Gro{\ss}mann}\ \emph {et~al.}(1997)\citenamefont
  {Gro{\ss}mann}, \citenamefont {Rost},\ and\ \citenamefont
  {Schleich}}]{Grossmann1997}%
  \BibitemOpen
  \bibfield  {author} {\bibinfo {author} {\bibfnamefont {F.}~\bibnamefont
  {Gro{\ss}mann}}, \bibinfo {author} {\bibfnamefont {J.-M.}\ \bibnamefont
  {Rost}},\ and\ \bibinfo {author} {\bibfnamefont {W.~P.}\ \bibnamefont
  {Schleich}},\ }\bibfield  {title} {\bibinfo {title} {{Spacetime structures in
  simple quantum systems}},\ }\href
  {https://doi.org/10.1088/0305-4470/30/9/004} {\bibfield  {journal} {\bibinfo
  {journal} {J. Phys. A. Math. Gen.}\ }\textbf {\bibinfo {volume} {30}},\
  \bibinfo {pages} {L277} (\bibinfo {year} {1997})}\BibitemShut {NoStop}%
\bibitem [{\citenamefont {Marzoli}\ \emph {et~al.}(1998)\citenamefont
  {Marzoli}, \citenamefont {Saif}, \citenamefont {Bialynicki-Birula},
  \citenamefont {Friesch}, \citenamefont {Kaplan},\ and\ \citenamefont
  {Schleich}}]{Marzoli1998}%
  \BibitemOpen
  \bibfield  {author} {\bibinfo {author} {\bibfnamefont {I.}~\bibnamefont
  {Marzoli}}, \bibinfo {author} {\bibfnamefont {F.}~\bibnamefont {Saif}},
  \bibinfo {author} {\bibfnamefont {I.}~\bibnamefont {Bialynicki-Birula}},
  \bibinfo {author} {\bibfnamefont {O.~M.}\ \bibnamefont {Friesch}}, \bibinfo
  {author} {\bibfnamefont {A.~E.}\ \bibnamefont {Kaplan}},\ and\ \bibinfo
  {author} {\bibfnamefont {W.~P.}\ \bibnamefont {Schleich}},\ }\bibfield
  {title} {\bibinfo {title} {{Quantum Carpets made simple}},\ }\href@noop {}
  {\bibfield  {journal} {\bibinfo  {journal} {Acta Phys. Slov.}\ }\textbf
  {\bibinfo {volume} {48}},\ \bibinfo {pages} {323} (\bibinfo {year}
  {1998})}\BibitemShut {NoStop}%
\bibitem [{\citenamefont {Kaplan}\ \emph {et~al.}(1998)\citenamefont {Kaplan},
  \citenamefont {Stifter}, \citenamefont {van Leeuwen}, \citenamefont {Lamb},\
  and\ \citenamefont {Schleich}}]{Kaplan1998a}%
  \BibitemOpen
  \bibfield  {author} {\bibinfo {author} {\bibfnamefont {A.~E.}\ \bibnamefont
  {Kaplan}}, \bibinfo {author} {\bibfnamefont {P.}~\bibnamefont {Stifter}},
  \bibinfo {author} {\bibfnamefont {K.~A.~H.}\ \bibnamefont {van Leeuwen}},
  \bibinfo {author} {\bibfnamefont {W.~E.}\ \bibnamefont {Lamb}},\ and\
  \bibinfo {author} {\bibfnamefont {W.~P.}\ \bibnamefont {Schleich}},\
  }\bibfield  {title} {\bibinfo {title} {{Intermode Traces -- Fundamental
  Interference Phenomenon in Quantum and Wave Physics}},\ }\href
  {https://doi.org/10.1238/Physica.Topical.076a00093} {\bibfield  {journal}
  {\bibinfo  {journal} {Phys. Scr.}\ }\textbf {\bibinfo {volume} {T76}},\
  \bibinfo {pages} {93} (\bibinfo {year} {1998})}\BibitemShut {NoStop}%
\bibitem [{\citenamefont {Berry}\ \emph {et~al.}(2001)\citenamefont {Berry},
  \citenamefont {Marzoli},\ and\ \citenamefont {Schleich}}]{Berry2001}%
  \BibitemOpen
  \bibfield  {author} {\bibinfo {author} {\bibfnamefont {M.}~\bibnamefont
  {Berry}}, \bibinfo {author} {\bibfnamefont {I.}~\bibnamefont {Marzoli}},\
  and\ \bibinfo {author} {\bibfnamefont {W.}~\bibnamefont {Schleich}},\
  }\bibfield  {title} {\bibinfo {title} {{Quantum carpets, carpets of light}},\
  }\href {https://doi.org/10.1088/2058-7058/14/6/30} {\bibfield  {journal}
  {\bibinfo  {journal} {Phys. World}\ }\textbf {\bibinfo {volume} {14}},\
  \bibinfo {pages} {39} (\bibinfo {year} {2001})}\BibitemShut {NoStop}%
\bibitem [{\citenamefont {Kazemi}\ \emph {et~al.}(2013)\citenamefont {Kazemi},
  \citenamefont {Chaturvedi}, \citenamefont {Marzoli}, \citenamefont
  {O'Connell},\ and\ \citenamefont {Schleich}}]{Kazemi2013}%
  \BibitemOpen
  \bibfield  {author} {\bibinfo {author} {\bibfnamefont {P.}~\bibnamefont
  {Kazemi}}, \bibinfo {author} {\bibfnamefont {S.}~\bibnamefont {Chaturvedi}},
  \bibinfo {author} {\bibfnamefont {I.}~\bibnamefont {Marzoli}}, \bibinfo
  {author} {\bibfnamefont {R.~F.}\ \bibnamefont {O'Connell}},\ and\ \bibinfo
  {author} {\bibfnamefont {W.~P.}\ \bibnamefont {Schleich}},\ }\bibfield
  {title} {\bibinfo {title} {{Quantum carpets: a tool to observe
  decoherence}},\ }\href {https://doi.org/10.1088/1367-2630/15/1/013052}
  {\bibfield  {journal} {\bibinfo  {journal} {New J. Phys.}\ }\textbf {\bibinfo
  {volume} {15}},\ \bibinfo {pages} {013052} (\bibinfo {year}
  {2013})}\BibitemShut {NoStop}%
\bibitem [{\citenamefont {Chen}\ \emph {et~al.}(2018)\citenamefont {Chen},
  \citenamefont {Beierle},\ and\ \citenamefont {Batelaan}}]{Chen2018}%
  \BibitemOpen
  \bibfield  {author} {\bibinfo {author} {\bibfnamefont {Z.}~\bibnamefont
  {Chen}}, \bibinfo {author} {\bibfnamefont {P.}~\bibnamefont {Beierle}},\ and\
  \bibinfo {author} {\bibfnamefont {H.}~\bibnamefont {Batelaan}},\ }\bibfield
  {title} {\bibinfo {title} {{Spatial correlation in matter-wave interference
  as a measure of decoherence, dephasing, and entropy}},\ }\href
  {https://doi.org/10.1103/PhysRevA.97.043608} {\bibfield  {journal} {\bibinfo
  {journal} {Phys. Rev. A}\ }\textbf {\bibinfo {volume} {97}},\ \bibinfo
  {pages} {043608} (\bibinfo {year} {2018})}\BibitemShut {NoStop}%
\bibitem [{\citenamefont {Friesch}\ \emph {et~al.}(2000)\citenamefont
  {Friesch}, \citenamefont {Marzoli},\ and\ \citenamefont
  {Schleich}}]{Friesch2000}%
  \BibitemOpen
  \bibfield  {author} {\bibinfo {author} {\bibfnamefont {O.~M.}\ \bibnamefont
  {Friesch}}, \bibinfo {author} {\bibfnamefont {I.}~\bibnamefont {Marzoli}},\
  and\ \bibinfo {author} {\bibfnamefont {W.~P.}\ \bibnamefont {Schleich}},\
  }\bibfield  {title} {\bibinfo {title} {{Quantum carpets woven by Wigner
  functions}},\ }\href {https://doi.org/10.1088/1367-2630/2/1/004} {\bibfield
  {journal} {\bibinfo  {journal} {New J. Phys.}\ }\textbf {\bibinfo {volume}
  {2}},\ \bibinfo {pages} {4} (\bibinfo {year} {2000})}\BibitemShut {NoStop}%
\bibitem [{\citenamefont {Loinaz}\ and\ \citenamefont
  {Newman}(1999)}]{Loinaz1999}%
  \BibitemOpen
  \bibfield  {author} {\bibinfo {author} {\bibfnamefont {W.}~\bibnamefont
  {Loinaz}}\ and\ \bibinfo {author} {\bibfnamefont {T.~J.}\ \bibnamefont
  {Newman}},\ }\bibfield  {title} {\bibinfo {title} {{Quantum revivals and
  carpets in some exactly solvable systems}},\ }\href
  {https://doi.org/10.1088/0305-4470/32/50/309} {\bibfield  {journal} {\bibinfo
   {journal} {J. Phys. A. Math. Gen.}\ }\textbf {\bibinfo {volume} {32}},\
  \bibinfo {pages} {8889} (\bibinfo {year} {1999})}\BibitemShut {NoStop}%
\bibitem [{\citenamefont {Hall}\ \emph {et~al.}(1999)\citenamefont {Hall},
  \citenamefont {Reineker},\ and\ \citenamefont {Schleich}}]{Hall1999}%
  \BibitemOpen
  \bibfield  {author} {\bibinfo {author} {\bibfnamefont {M.~J.~W.}\
  \bibnamefont {Hall}}, \bibinfo {author} {\bibfnamefont {M.~S.}\ \bibnamefont
  {Reineker}},\ and\ \bibinfo {author} {\bibfnamefont {W.~P.}\ \bibnamefont
  {Schleich}},\ }\bibfield  {title} {\bibinfo {title} {{Unravelling quantum
  carpets: a travelling-wave approach}},\ }\href
  {https://doi.org/10.1088/0305-4470/32/47/307} {\bibfield  {journal} {\bibinfo
   {journal} {J. Phys. A. Math. Gen.}\ }\textbf {\bibinfo {volume} {32}},\
  \bibinfo {pages} {8275} (\bibinfo {year} {1999})}\BibitemShut {NoStop}%
\bibitem [{\citenamefont {Banchi}\ \emph {et~al.}(2015)\citenamefont {Banchi},
  \citenamefont {Compagno},\ and\ \citenamefont {Bose}}]{Banchi2015}%
  \BibitemOpen
  \bibfield  {author} {\bibinfo {author} {\bibfnamefont {L.}~\bibnamefont
  {Banchi}}, \bibinfo {author} {\bibfnamefont {E.}~\bibnamefont {Compagno}},\
  and\ \bibinfo {author} {\bibfnamefont {S.}~\bibnamefont {Bose}},\ }\bibfield
  {title} {\bibinfo {title} {{Perfect wave-packet splitting and reconstruction
  in a one-dimensional lattice}},\ }\href
  {https://doi.org/10.1103/PhysRevA.91.052323} {\bibfield  {journal} {\bibinfo
  {journal} {Phys. Rev. A}\ }\textbf {\bibinfo {volume} {91}},\ \bibinfo
  {pages} {052323} (\bibinfo {year} {2015})}\BibitemShut {NoStop}%
\bibitem [{\citenamefont {Genest}\ \emph
  {et~al.}(2016{\natexlab{a}})\citenamefont {Genest}, \citenamefont {Vinet},\
  and\ \citenamefont {Zhedanov}}]{Genest2016a}%
  \BibitemOpen
  \bibfield  {author} {\bibinfo {author} {\bibfnamefont {V.~X.}\ \bibnamefont
  {Genest}}, \bibinfo {author} {\bibfnamefont {L.}~\bibnamefont {Vinet}},\ and\
  \bibinfo {author} {\bibfnamefont {A.}~\bibnamefont {Zhedanov}},\ }\bibfield
  {title} {\bibinfo {title} {{Quantum spin chains with fractional revival}},\
  }\href {https://doi.org/10.1016/J.AOP.2016.05.009} {\bibfield  {journal}
  {\bibinfo  {journal} {Ann. Phys. (N. Y).}\ }\textbf {\bibinfo {volume}
  {371}},\ \bibinfo {pages} {348} (\bibinfo {year}
  {2016}{\natexlab{a}})}\BibitemShut {NoStop}%
\bibitem [{\citenamefont {Genest}\ \emph
  {et~al.}(2016{\natexlab{b}})\citenamefont {Genest}, \citenamefont {Vinet},\
  and\ \citenamefont {Zhedanov}}]{Genest2016}%
  \BibitemOpen
  \bibfield  {author} {\bibinfo {author} {\bibfnamefont {V.~X.}\ \bibnamefont
  {Genest}}, \bibinfo {author} {\bibfnamefont {L.}~\bibnamefont {Vinet}},\ and\
  \bibinfo {author} {\bibfnamefont {A.}~\bibnamefont {Zhedanov}},\ }\bibfield
  {title} {\bibinfo {title} {{Exact fractional revival in spin chains}},\
  }\href {https://doi.org/10.1142/S0217984916503152} {\bibfield  {journal}
  {\bibinfo  {journal} {Mod. Phys. Lett. B}\ }\textbf {\bibinfo {volume}
  {30}},\ \bibinfo {pages} {1650315} (\bibinfo {year}
  {2016}{\natexlab{b}})}\BibitemShut {NoStop}%
\bibitem [{\citenamefont {Lemay}\ \emph {et~al.}(2016)\citenamefont {Lemay},
  \citenamefont {Vinet},\ and\ \citenamefont {Zhedanov}}]{Lemay2016}%
  \BibitemOpen
  \bibfield  {author} {\bibinfo {author} {\bibfnamefont {J.-M.}\ \bibnamefont
  {Lemay}}, \bibinfo {author} {\bibfnamefont {L.}~\bibnamefont {Vinet}},\ and\
  \bibinfo {author} {\bibfnamefont {A.}~\bibnamefont {Zhedanov}},\ }\bibfield
  {title} {\bibinfo {title} {{An analytic spin chain model with fractional
  revival}},\ }\href {https://doi.org/10.1088/1751-8113/49/33/335302}
  {\bibfield  {journal} {\bibinfo  {journal} {J. Phys. A Math. Theor.}\
  }\textbf {\bibinfo {volume} {49}},\ \bibinfo {pages} {335302} (\bibinfo
  {year} {2016})}\BibitemShut {NoStop}%
\bibitem [{\citenamefont {Kay}(2017)}]{Kay2017}%
  \BibitemOpen
  \bibfield  {author} {\bibinfo {author} {\bibfnamefont {A.}~\bibnamefont
  {Kay}},\ }\bibfield  {title} {\bibinfo {title} {{Generating Quantum States
  through Spin Chain Dynamics}},\ }\href
  {https://doi.org/10.1088/1367-2630/aa68f9} {\bibfield  {journal} {\bibinfo
  {journal} {New J. Phys.}\ }\textbf {\bibinfo {volume} {19}},\ \bibinfo
  {pages} {043019} (\bibinfo {year} {2017})}\BibitemShut {NoStop}%
\bibitem [{\citenamefont {Compagno}\ \emph {et~al.}(2017)\citenamefont
  {Compagno}, \citenamefont {Banchi}, \citenamefont {Gross},\ and\
  \citenamefont {Bose}}]{Compagno2016}%
  \BibitemOpen
  \bibfield  {author} {\bibinfo {author} {\bibfnamefont {E.}~\bibnamefont
  {Compagno}}, \bibinfo {author} {\bibfnamefont {L.}~\bibnamefont {Banchi}},
  \bibinfo {author} {\bibfnamefont {C.}~\bibnamefont {Gross}},\ and\ \bibinfo
  {author} {\bibfnamefont {S.}~\bibnamefont {Bose}},\ }\bibfield  {title}
  {\bibinfo {title} {{NOON States via Quantum Walk of Bound Particles}},\
  }\href {https://doi.org/10.1103/PhysRevA.95.012307} {\bibfield  {journal}
  {\bibinfo  {journal} {Phys. Rev. A}\ }\textbf {\bibinfo {volume} {95}},\
  \bibinfo {pages} {012307} (\bibinfo {year} {2017})}\BibitemShut {NoStop}%
\bibitem [{\citenamefont {Christandl}\ \emph {et~al.}(2017)\citenamefont
  {Christandl}, \citenamefont {Vinet},\ and\ \citenamefont
  {Zhedanov}}]{Christandl2017}%
  \BibitemOpen
  \bibfield  {author} {\bibinfo {author} {\bibfnamefont {M.}~\bibnamefont
  {Christandl}}, \bibinfo {author} {\bibfnamefont {L.}~\bibnamefont {Vinet}},\
  and\ \bibinfo {author} {\bibfnamefont {A.}~\bibnamefont {Zhedanov}},\
  }\bibfield  {title} {\bibinfo {title} {{Analytic next-to-nearest-neighbor X X
  models with perfect state transfer and fractional revival}},\ }\href
  {https://doi.org/10.1103/PhysRevA.96.032335} {\bibfield  {journal} {\bibinfo
  {journal} {Phys. Rev. A}\ }\textbf {\bibinfo {volume} {96}},\ \bibinfo
  {pages} {032335} (\bibinfo {year} {2017})}\BibitemShut {NoStop}%
\bibitem [{\citenamefont {Aronstein}\ and\ \citenamefont
  {Stroud}(1997)}]{Aronstein1997}%
  \BibitemOpen
  \bibfield  {author} {\bibinfo {author} {\bibfnamefont {D.~L.}\ \bibnamefont
  {Aronstein}}\ and\ \bibinfo {author} {\bibfnamefont {C.~R.}\ \bibnamefont
  {Stroud}},\ }\bibfield  {title} {\bibinfo {title} {{Fractional wave-function
  revivals in the infinite square well}},\ }\href
  {https://doi.org/10.1103/PhysRevA.55.4526} {\bibfield  {journal} {\bibinfo
  {journal} {Phys. Rev. A}\ }\textbf {\bibinfo {volume} {55}},\ \bibinfo
  {pages} {4526} (\bibinfo {year} {1997})}\BibitemShut {NoStop}%
\bibitem [{\citenamefont {Nest}(2006)}]{Nest2006}%
  \BibitemOpen
  \bibfield  {author} {\bibinfo {author} {\bibfnamefont {M.}~\bibnamefont
  {Nest}},\ }\bibfield  {title} {\bibinfo {title} {{Quantum carpets and
  correlated dynamics of several fermions}},\ }\href
  {https://doi.org/10.1103/PhysRevA.73.023613} {\bibfield  {journal} {\bibinfo
  {journal} {Phys. Rev. A}\ }\textbf {\bibinfo {volume} {73}},\ \bibinfo
  {pages} {023613} (\bibinfo {year} {2006})}\BibitemShut {NoStop}%
\bibitem [{\citenamefont {Ruostekoski}\ \emph {et~al.}(2001)\citenamefont
  {Ruostekoski}, \citenamefont {Kneer}, \citenamefont {Schleich},\ and\
  \citenamefont {Rempe}}]{Ruostekoski2001}%
  \BibitemOpen
  \bibfield  {author} {\bibinfo {author} {\bibfnamefont {J.}~\bibnamefont
  {Ruostekoski}}, \bibinfo {author} {\bibfnamefont {B.}~\bibnamefont {Kneer}},
  \bibinfo {author} {\bibfnamefont {W.~P.}\ \bibnamefont {Schleich}},\ and\
  \bibinfo {author} {\bibfnamefont {G.}~\bibnamefont {Rempe}},\ }\bibfield
  {title} {\bibinfo {title} {{Interference of a Bose-Einstein condensate in a
  hard-wall trap: From the nonlinear Talbot effect to the formation of
  vorticity}},\ }\href {https://doi.org/10.1103/PhysRevA.63.043613} {\bibfield
  {journal} {\bibinfo  {journal} {Phys. Rev. A}\ }\textbf {\bibinfo {volume}
  {63}},\ \bibinfo {pages} {043613} (\bibinfo {year} {2001})}\BibitemShut
  {NoStop}%
\bibitem [{\citenamefont {Gawryluk}\ \emph {et~al.}(2006)\citenamefont
  {Gawryluk}, \citenamefont {Brewczyk}, \citenamefont {Gajda},\ and\
  \citenamefont {Mostowski}}]{Gawryluk2006}%
  \BibitemOpen
  \bibfield  {author} {\bibinfo {author} {\bibfnamefont {K.}~\bibnamefont
  {Gawryluk}}, \bibinfo {author} {\bibfnamefont {M.}~\bibnamefont {Brewczyk}},
  \bibinfo {author} {\bibfnamefont {M.}~\bibnamefont {Gajda}},\ and\ \bibinfo
  {author} {\bibfnamefont {J.}~\bibnamefont {Mostowski}},\ }\bibfield  {title}
  {\bibinfo {title} {{Formation of soliton trains in Bose–Einstein
  condensates by temporal Talbot effect}},\ }\href
  {https://doi.org/10.1088/0953-4075/39/1/L01} {\bibfield  {journal} {\bibinfo
  {journal} {J. Phys. B At. Mol. Opt. Phys.}\ }\textbf {\bibinfo {volume}
  {39}},\ \bibinfo {pages} {L1} (\bibinfo {year} {2006})}\BibitemShut {NoStop}%
\bibitem [{\citenamefont {Grochowski}\ \emph {et~al.}(2020)\citenamefont
  {Grochowski}, \citenamefont {Karpiuk}, \citenamefont {Brewczyk},\ and\
  \citenamefont {Rz\k{a}\ifmmode~\dot{z}\else
  \.{z}\fi{}ewski}}]{Grochowski2020}%
  \BibitemOpen
  \bibfield  {author} {\bibinfo {author} {\bibfnamefont {P.~T.}\ \bibnamefont
  {Grochowski}}, \bibinfo {author} {\bibfnamefont {T.}~\bibnamefont {Karpiuk}},
  \bibinfo {author} {\bibfnamefont {M.}~\bibnamefont {Brewczyk}},\ and\
  \bibinfo {author} {\bibfnamefont {K.}~\bibnamefont
  {Rz\k{a}\ifmmode~\dot{z}\else \.{z}\fi{}ewski}},\ }\bibfield  {title}
  {\bibinfo {title} {{Fermionic quantum carpets: From canals and ridges to
  solitonlike structures}},\ }\href
  {https://doi.org/10.1103/PhysRevResearch.2.013119} {\bibfield  {journal}
  {\bibinfo  {journal} {Phys. Rev. Research}\ }\textbf {\bibinfo {volume}
  {2}},\ \bibinfo {pages} {013119} (\bibinfo {year} {2020})}\BibitemShut
  {NoStop}%
\bibitem [{\citenamefont {Nowak}\ \emph {et~al.}(1997)\citenamefont {Nowak},
  \citenamefont {Kurtsiefer}, \citenamefont {Pfau},\ and\ \citenamefont
  {David}}]{Nowak1997}%
  \BibitemOpen
  \bibfield  {author} {\bibinfo {author} {\bibfnamefont {S.}~\bibnamefont
  {Nowak}}, \bibinfo {author} {\bibfnamefont {C.}~\bibnamefont {Kurtsiefer}},
  \bibinfo {author} {\bibfnamefont {T.}~\bibnamefont {Pfau}},\ and\ \bibinfo
  {author} {\bibfnamefont {C.}~\bibnamefont {David}},\ }\bibfield  {title}
  {\bibinfo {title} {{High-order Talbot fringes for atomic matter waves}},\
  }\href {https://doi.org/10.1364/OL.22.001430} {\bibfield  {journal} {\bibinfo
   {journal} {Opt. Lett.}\ }\textbf {\bibinfo {volume} {22}},\ \bibinfo {pages}
  {1430} (\bibinfo {year} {1997})}\BibitemShut {NoStop}%
\bibitem [{\citenamefont {Chapman}\ \emph
  {et~al.}(1995{\natexlab{a}})\citenamefont {Chapman}, \citenamefont {Ekstrom},
  \citenamefont {Hammond}, \citenamefont {Schmiedmayer}, \citenamefont
  {Tannian}, \citenamefont {Wehinger},\ and\ \citenamefont
  {Pritchard}}]{Chapman1995}%
  \BibitemOpen
  \bibfield  {author} {\bibinfo {author} {\bibfnamefont {M.~S.}\ \bibnamefont
  {Chapman}}, \bibinfo {author} {\bibfnamefont {C.~R.}\ \bibnamefont
  {Ekstrom}}, \bibinfo {author} {\bibfnamefont {T.~D.}\ \bibnamefont
  {Hammond}}, \bibinfo {author} {\bibfnamefont {J.}~\bibnamefont
  {Schmiedmayer}}, \bibinfo {author} {\bibfnamefont {B.~E.}\ \bibnamefont
  {Tannian}}, \bibinfo {author} {\bibfnamefont {S.}~\bibnamefont {Wehinger}},\
  and\ \bibinfo {author} {\bibfnamefont {D.~E.}\ \bibnamefont {Pritchard}},\
  }\bibfield  {title} {\bibinfo {title} {{Near-field imaging of atom
  diffraction gratings: The atomic Talbot effect}},\ }\href
  {https://doi.org/10.1103/PhysRevA.51.R14} {\bibfield  {journal} {\bibinfo
  {journal} {Phys. Rev. A}\ }\textbf {\bibinfo {volume} {51}},\ \bibinfo
  {pages} {R14} (\bibinfo {year} {1995}{\natexlab{a}})}\BibitemShut {NoStop}%
\bibitem [{\citenamefont {Ryu}\ \emph {et~al.}(2006)\citenamefont {Ryu},
  \citenamefont {Andersen}, \citenamefont {Vaziri}, \citenamefont {D'Arcy},
  \citenamefont {Grossman}, \citenamefont {Helmerson},\ and\ \citenamefont
  {Phillips}}]{Ryu2006}%
  \BibitemOpen
  \bibfield  {author} {\bibinfo {author} {\bibfnamefont {C.}~\bibnamefont
  {Ryu}}, \bibinfo {author} {\bibfnamefont {M.~F.}\ \bibnamefont {Andersen}},
  \bibinfo {author} {\bibfnamefont {A.}~\bibnamefont {Vaziri}}, \bibinfo
  {author} {\bibfnamefont {M.~B.}\ \bibnamefont {D'Arcy}}, \bibinfo {author}
  {\bibfnamefont {J.~M.}\ \bibnamefont {Grossman}}, \bibinfo {author}
  {\bibfnamefont {K.}~\bibnamefont {Helmerson}},\ and\ \bibinfo {author}
  {\bibfnamefont {W.~D.}\ \bibnamefont {Phillips}},\ }\bibfield  {title}
  {\bibinfo {title} {{High-Order Quantum Resonances Observed in a Periodically
  Kicked Bose-Einstein Condensate}},\ }\href
  {https://doi.org/10.1103/PhysRevLett.96.160403} {\bibfield  {journal}
  {\bibinfo  {journal} {Phys. Rev. Lett.}\ }\textbf {\bibinfo {volume} {96}},\
  \bibinfo {pages} {160403} (\bibinfo {year} {2006})}\BibitemShut {NoStop}%
\bibitem [{\citenamefont {Deng}\ \emph {et~al.}(1999)\citenamefont {Deng},
  \citenamefont {Hagley}, \citenamefont {Denschlag}, \citenamefont {Simsarian},
  \citenamefont {Edwards}, \citenamefont {Clark}, \citenamefont {Helmerson},
  \citenamefont {Rolston},\ and\ \citenamefont {Phillips}}]{Deng1999}%
  \BibitemOpen
  \bibfield  {author} {\bibinfo {author} {\bibfnamefont {L.}~\bibnamefont
  {Deng}}, \bibinfo {author} {\bibfnamefont {E.~W.}\ \bibnamefont {Hagley}},
  \bibinfo {author} {\bibfnamefont {J.}~\bibnamefont {Denschlag}}, \bibinfo
  {author} {\bibfnamefont {J.~E.}\ \bibnamefont {Simsarian}}, \bibinfo {author}
  {\bibfnamefont {M.}~\bibnamefont {Edwards}}, \bibinfo {author} {\bibfnamefont
  {C.~W.}\ \bibnamefont {Clark}}, \bibinfo {author} {\bibfnamefont
  {K.}~\bibnamefont {Helmerson}}, \bibinfo {author} {\bibfnamefont {S.~L.}\
  \bibnamefont {Rolston}},\ and\ \bibinfo {author} {\bibfnamefont {W.~D.}\
  \bibnamefont {Phillips}},\ }\bibfield  {title} {\bibinfo {title} {{Temporal,
  Matter-Wave-Dispersion Talbot Effect}},\ }\href
  {https://doi.org/10.1103/PhysRevLett.83.5407} {\bibfield  {journal} {\bibinfo
   {journal} {Phys. Rev. Lett.}\ }\textbf {\bibinfo {volume} {83}},\ \bibinfo
  {pages} {5407} (\bibinfo {year} {1999})}\BibitemShut {NoStop}%
\bibitem [{\citenamefont {Mark}\ \emph {et~al.}(2011)\citenamefont {Mark},
  \citenamefont {Haller}, \citenamefont {Danzl}, \citenamefont {Lauber},
  \citenamefont {Gustavsson},\ and\ \citenamefont {N{\"{a}}gerl}}]{Mark2011}%
  \BibitemOpen
  \bibfield  {author} {\bibinfo {author} {\bibfnamefont {M.~J.}\ \bibnamefont
  {Mark}}, \bibinfo {author} {\bibfnamefont {E.}~\bibnamefont {Haller}},
  \bibinfo {author} {\bibfnamefont {J.~G.}\ \bibnamefont {Danzl}}, \bibinfo
  {author} {\bibfnamefont {K.}~\bibnamefont {Lauber}}, \bibinfo {author}
  {\bibfnamefont {M.}~\bibnamefont {Gustavsson}},\ and\ \bibinfo {author}
  {\bibfnamefont {H.-C.}\ \bibnamefont {N{\"{a}}gerl}},\ }\bibfield  {title}
  {\bibinfo {title} {{Demonstration of the temporal matter-wave Talbot effect
  for trapped matter waves}},\ }\href
  {https://doi.org/10.1088/1367-2630/13/8/085008} {\bibfield  {journal}
  {\bibinfo  {journal} {New J. Phys.}\ }\textbf {\bibinfo {volume} {13}},\
  \bibinfo {pages} {085008} (\bibinfo {year} {2011})}\BibitemShut {NoStop}%
\bibitem [{\citenamefont {Ahn}\ \emph {et~al.}(2001)\citenamefont {Ahn},
  \citenamefont {Hutchinson}, \citenamefont {Rangan},\ and\ \citenamefont
  {Bucksbaum}}]{Ahn2001}%
  \BibitemOpen
  \bibfield  {author} {\bibinfo {author} {\bibfnamefont {J.}~\bibnamefont
  {Ahn}}, \bibinfo {author} {\bibfnamefont {D.~N.}\ \bibnamefont {Hutchinson}},
  \bibinfo {author} {\bibfnamefont {C.}~\bibnamefont {Rangan}},\ and\ \bibinfo
  {author} {\bibfnamefont {P.~H.}\ \bibnamefont {Bucksbaum}},\ }\bibfield
  {title} {\bibinfo {title} {{Quantum Phase Retrieval of a Rydberg Wave Packet
  Using a Half-Cycle Pulse}},\ }\href
  {https://doi.org/10.1103/PhysRevLett.86.1179} {\bibfield  {journal} {\bibinfo
   {journal} {Phys. Rev. Lett.}\ }\textbf {\bibinfo {volume} {86}},\ \bibinfo
  {pages} {1179} (\bibinfo {year} {2001})}\BibitemShut {NoStop}%
\bibitem [{\citenamefont {Katsuki}\ \emph {et~al.}(2009)\citenamefont
  {Katsuki}, \citenamefont {Chiba}, \citenamefont {Meier}, \citenamefont
  {Girard},\ and\ \citenamefont {Ohmori}}]{Katsuki2009}%
  \BibitemOpen
  \bibfield  {author} {\bibinfo {author} {\bibfnamefont {H.}~\bibnamefont
  {Katsuki}}, \bibinfo {author} {\bibfnamefont {H.}~\bibnamefont {Chiba}},
  \bibinfo {author} {\bibfnamefont {C.}~\bibnamefont {Meier}}, \bibinfo
  {author} {\bibfnamefont {B.}~\bibnamefont {Girard}},\ and\ \bibinfo {author}
  {\bibfnamefont {K.}~\bibnamefont {Ohmori}},\ }\bibfield  {title} {\bibinfo
  {title} {{Actively Tailored Spatiotemporal Images of Quantum Interference on
  the Picometer and Femtosecond Scales}},\ }\href
  {https://doi.org/10.1103/PhysRevLett.102.103602} {\bibfield  {journal}
  {\bibinfo  {journal} {Phys. Rev. Lett.}\ }\textbf {\bibinfo {volume} {102}},\
  \bibinfo {pages} {103602} (\bibinfo {year} {2009})}\BibitemShut {NoStop}%
\bibitem [{\citenamefont {Bethe}(1931)}]{Bethe1931}%
  \BibitemOpen
  \bibfield  {author} {\bibinfo {author} {\bibfnamefont {H.~A.}\ \bibnamefont
  {Bethe}},\ }\bibfield  {title} {\bibinfo {title} {{Zur Theorie der
  Metalle}},\ }\href@noop {} {\bibfield  {journal} {\bibinfo  {journal}
  {Zeitschrift f{\"{u}}r Phys.}\ }\textbf {\bibinfo {volume} {71}},\ \bibinfo
  {pages} {205} (\bibinfo {year} {1931})}\BibitemShut {NoStop}%
\bibitem [{\citenamefont {Lieb}\ and\ \citenamefont
  {Liniger}(1963)}]{Lieb1963b}%
  \BibitemOpen
  \bibfield  {author} {\bibinfo {author} {\bibfnamefont {E.~H.}\ \bibnamefont
  {Lieb}}\ and\ \bibinfo {author} {\bibfnamefont {W.}~\bibnamefont {Liniger}},\
  }\bibfield  {title} {\bibinfo {title} {{Exact analysis of an interacting bose
  gas. I. the general solution and the ground state}},\ }\href
  {https://doi.org/10.1103/PhysRev.130.1605} {\bibfield  {journal} {\bibinfo
  {journal} {Phys. Rev.}\ }\textbf {\bibinfo {volume} {130}},\ \bibinfo {pages}
  {1605} (\bibinfo {year} {1963})}\BibitemShut {NoStop}%
\bibitem [{\citenamefont {Gaudin}(1971)}]{Gaudin1971}%
  \BibitemOpen
  \bibfield  {author} {\bibinfo {author} {\bibfnamefont {M.}~\bibnamefont
  {Gaudin}},\ }\bibfield  {title} {\bibinfo {title} {{Boundary energy of a bose
  gas in one dimension}},\ }\href {https://doi.org/10.1103/PhysRevA.4.386}
  {\bibfield  {journal} {\bibinfo  {journal} {Phys. Rev. A}\ }\textbf {\bibinfo
  {volume} {4}},\ \bibinfo {pages} {386} (\bibinfo {year} {1971})}\BibitemShut
  {NoStop}%
\bibitem [{\citenamefont {Batchelor}\ \emph {et~al.}(2005)\citenamefont
  {Batchelor}, \citenamefont {Guan}, \citenamefont {Oelkers},\ and\
  \citenamefont {Lee}}]{Batchelor2005}%
  \BibitemOpen
  \bibfield  {author} {\bibinfo {author} {\bibfnamefont {M.~T.}\ \bibnamefont
  {Batchelor}}, \bibinfo {author} {\bibfnamefont {X.~W.}\ \bibnamefont {Guan}},
  \bibinfo {author} {\bibfnamefont {N.}~\bibnamefont {Oelkers}},\ and\ \bibinfo
  {author} {\bibfnamefont {C.}~\bibnamefont {Lee}},\ }\bibfield  {title}
  {\bibinfo {title} {{The ID interacting Bose gas in a hard wall box}},\ }\href
  {https://doi.org/10.1088/0305-4470/38/36/001} {\bibfield  {journal} {\bibinfo
   {journal} {J. Phys. A. Math. Gen.}\ }\textbf {\bibinfo {volume} {38}},\
  \bibinfo {pages} {7787} (\bibinfo {year} {2005})}\BibitemShut {NoStop}%
\bibitem [{\citenamefont {Oelkers}\ \emph {et~al.}(2006)\citenamefont
  {Oelkers}, \citenamefont {Batchelor}, \citenamefont {Bortz},\ and\
  \citenamefont {Guan}}]{Oelkers2006a}%
  \BibitemOpen
  \bibfield  {author} {\bibinfo {author} {\bibfnamefont {N.}~\bibnamefont
  {Oelkers}}, \bibinfo {author} {\bibfnamefont {M.~T.}\ \bibnamefont
  {Batchelor}}, \bibinfo {author} {\bibfnamefont {M.}~\bibnamefont {Bortz}},\
  and\ \bibinfo {author} {\bibfnamefont {X.~W.}\ \bibnamefont {Guan}},\
  }\bibfield  {title} {\bibinfo {title} {{Bethe ansatz study of one-dimensional
  Bose and Fermi gases with periodic and hard wall boundary conditions}},\
  }\href {https://doi.org/10.1088/0305-4470/39/5/005} {\bibfield  {journal}
  {\bibinfo  {journal} {J. Phys. A. Math. Gen.}\ }\textbf {\bibinfo {volume}
  {39}},\ \bibinfo {pages} {1073} (\bibinfo {year} {2006})}\BibitemShut
  {NoStop}%
\bibitem [{\citenamefont {Gaudin}\ and\ \citenamefont
  {Caux}(2012)}]{Gaudin2012}%
  \BibitemOpen
  \bibfield  {author} {\bibinfo {author} {\bibfnamefont {M.}~\bibnamefont
  {Gaudin}}\ and\ \bibinfo {author} {\bibfnamefont {J.~S.}\ \bibnamefont
  {Caux}},\ }\href {https://doi.org/10.1017/CBO9781107053885} {\emph {\bibinfo
  {title} {The Bethe Wavefunction}}}\ (\bibinfo  {publisher} {Cambridge
  University Press},\ \bibinfo {year} {2012})\BibitemShut {NoStop}%
\bibitem [{\citenamefont {Tomchenko}(2015)}]{Tomchenko2015}%
  \BibitemOpen
  \bibfield  {author} {\bibinfo {author} {\bibfnamefont {M.}~\bibnamefont
  {Tomchenko}},\ }\bibfield  {title} {\bibinfo {title} {{Point bosons in a
  one-dimensional box: The ground state, excitations and thermodynamics}},\
  }\href {https://doi.org/10.1088/1751-8113/48/36/365003} {\bibfield  {journal}
  {\bibinfo  {journal} {J. Phys. A Math. Theor.}\ }\textbf {\bibinfo {volume}
  {48}},\ \bibinfo {pages} {365003} (\bibinfo {year} {2015})}\BibitemShut
  {NoStop}%
\bibitem [{\citenamefont {Syrwid}\ \emph {et~al.}(2016)\citenamefont {Syrwid},
  \citenamefont {Brewczyk}, \citenamefont {Gajda},\ and\ \citenamefont
  {Sacha}}]{Syrwid2016}%
  \BibitemOpen
  \bibfield  {author} {\bibinfo {author} {\bibfnamefont {A.}~\bibnamefont
  {Syrwid}}, \bibinfo {author} {\bibfnamefont {M.}~\bibnamefont {Brewczyk}},
  \bibinfo {author} {\bibfnamefont {M.}~\bibnamefont {Gajda}},\ and\ \bibinfo
  {author} {\bibfnamefont {K.}~\bibnamefont {Sacha}},\ }\bibfield  {title}
  {\bibinfo {title} {{Single-shot simulations of dynamics of quantum dark
  solitons}},\ }\href {https://doi.org/10.1103/PhysRevA.94.023623} {\bibfield
  {journal} {\bibinfo  {journal} {Phys. Rev. A}\ }\textbf {\bibinfo {volume}
  {94}},\ \bibinfo {pages} {023623} (\bibinfo {year} {2016})}\BibitemShut
  {NoStop}%
\bibitem [{\citenamefont {Staro\ifmmode~\acute{n}\else \'{n}\fi{}}\ \emph
  {et~al.}(2020)\citenamefont {Staro\ifmmode~\acute{n}\else \'{n}\fi{}},
  \citenamefont {Syrwid},\ and\ \citenamefont {Sacha}}]{Staron2020}%
  \BibitemOpen
  \bibfield  {author} {\bibinfo {author} {\bibfnamefont {P.}~\bibnamefont
  {Staro\ifmmode~\acute{n}\else \'{n}\fi{}}}, \bibinfo {author} {\bibfnamefont
  {A.}~\bibnamefont {Syrwid}},\ and\ \bibinfo {author} {\bibfnamefont
  {K.}~\bibnamefont {Sacha}},\ }\bibfield  {title} {\bibinfo {title}
  {Measurement of a one-dimensional matter-wave quantum breather},\ }\href
  {https://doi.org/10.1103/PhysRevA.102.063308} {\bibfield  {journal} {\bibinfo
   {journal} {Phys. Rev. A}\ }\textbf {\bibinfo {volume} {102}},\ \bibinfo
  {pages} {063308} (\bibinfo {year} {2020})}\BibitemShut {NoStop}%
\bibitem [{\citenamefont {Syrwid}(2020)}]{Syrwid2020}%
  \BibitemOpen
  \bibfield  {author} {\bibinfo {author} {\bibfnamefont {A.}~\bibnamefont
  {Syrwid}},\ }\bibfield  {title} {\bibinfo {title} {Quantum dark solitons in
  ultracold one-dimensional bose and fermi gases},\ }\bibfield  {journal}
  {\bibinfo  {journal} {J. Phys. B}\ }\href
  {https://doi.org/doi:10.1088/1361-6455/abd37f} {doi:10.1088/1361-6455/abd37f}
  (\bibinfo {year} {2020})\BibitemShut {NoStop}%
\bibitem [{\citenamefont {Wilson}\ \emph {et~al.}(2020)\citenamefont {Wilson},
  \citenamefont {Malvania}, \citenamefont {Le}, \citenamefont {Zhang},
  \citenamefont {Rigol},\ and\ \citenamefont {Weiss}}]{Wilson2020}%
  \BibitemOpen
  \bibfield  {author} {\bibinfo {author} {\bibfnamefont {J.~M.}\ \bibnamefont
  {Wilson}}, \bibinfo {author} {\bibfnamefont {N.}~\bibnamefont {Malvania}},
  \bibinfo {author} {\bibfnamefont {Y.}~\bibnamefont {Le}}, \bibinfo {author}
  {\bibfnamefont {Y.}~\bibnamefont {Zhang}}, \bibinfo {author} {\bibfnamefont
  {M.}~\bibnamefont {Rigol}},\ and\ \bibinfo {author} {\bibfnamefont {D.~S.}\
  \bibnamefont {Weiss}},\ }\bibfield  {title} {\bibinfo {title} {Observation of
  dynamical fermionization},\ }\href {https://doi.org/10.1126/science.aaz0242}
  {\bibfield  {journal} {\bibinfo  {journal} {Science}\ }\textbf {\bibinfo
  {volume} {367}},\ \bibinfo {pages} {1461} (\bibinfo {year}
  {2020})}\BibitemShut {NoStop}%
\bibitem [{\citenamefont {Girardeau}(1960)}]{Girardeau1960a}%
  \BibitemOpen
  \bibfield  {author} {\bibinfo {author} {\bibfnamefont {M.}~\bibnamefont
  {Girardeau}},\ }\bibfield  {title} {\bibinfo {title} {{Relationship between
  systems of impenetrable bosons and fermions in one dimension}},\ }\href
  {https://doi.org/10.1063/1.1703687} {\bibfield  {journal} {\bibinfo
  {journal} {J. Math. Phys.}\ }\textbf {\bibinfo {volume} {1}},\ \bibinfo
  {pages} {516} (\bibinfo {year} {1960})}\BibitemShut {NoStop}%
\bibitem [{\citenamefont {Yukalov}\ and\ \citenamefont
  {Girardeau}(2005)}]{Yukalov2005}%
  \BibitemOpen
  \bibfield  {author} {\bibinfo {author} {\bibfnamefont {V.~I.}\ \bibnamefont
  {Yukalov}}\ and\ \bibinfo {author} {\bibfnamefont {M.~D.}\ \bibnamefont
  {Girardeau}},\ }\bibfield  {title} {\bibinfo {title} {{Fermi-Bose mapping for
  one-dimensional Bose gases}},\ }\href
  {https://doi.org/10.1002/lapl.200510011} {\bibfield  {journal} {\bibinfo
  {journal} {Laser Phys. Lett.}\ }\textbf {\bibinfo {volume} {2}},\ \bibinfo
  {pages} {375} (\bibinfo {year} {2005})}\BibitemShut {NoStop}%
\bibitem [{\citenamefont {Rigol}\ and\ \citenamefont
  {Muramatsu}(2005)}]{Rigol2005}%
  \BibitemOpen
  \bibfield  {author} {\bibinfo {author} {\bibfnamefont {M.}~\bibnamefont
  {Rigol}}\ and\ \bibinfo {author} {\bibfnamefont {A.}~\bibnamefont
  {Muramatsu}},\ }\bibfield  {title} {\bibinfo {title} {Fermionization in an
  expanding 1d gas of hard-core bosons},\ }\href
  {https://doi.org/10.1103/PhysRevLett.94.240403} {\bibfield  {journal}
  {\bibinfo  {journal} {Phys. Rev. Lett.}\ }\textbf {\bibinfo {volume} {94}},\
  \bibinfo {pages} {240403} (\bibinfo {year} {2005})}\BibitemShut {NoStop}%
\bibitem [{\citenamefont {Minguzzi}\ and\ \citenamefont
  {Gangardt}(2005)}]{Minguzzi2005}%
  \BibitemOpen
  \bibfield  {author} {\bibinfo {author} {\bibfnamefont {A.}~\bibnamefont
  {Minguzzi}}\ and\ \bibinfo {author} {\bibfnamefont {D.~M.}\ \bibnamefont
  {Gangardt}},\ }\bibfield  {title} {\bibinfo {title} {Exact coherent states of
  a harmonically confined tonks-girardeau gas},\ }\href
  {https://doi.org/10.1103/PhysRevLett.94.240404} {\bibfield  {journal}
  {\bibinfo  {journal} {Phys. Rev. Lett.}\ }\textbf {\bibinfo {volume} {94}},\
  \bibinfo {pages} {240404} (\bibinfo {year} {2005})}\BibitemShut {NoStop}%
\bibitem [{\citenamefont {Pezer}\ and\ \citenamefont
  {Buljan}(2007)}]{Pezer2007a}%
  \BibitemOpen
  \bibfield  {author} {\bibinfo {author} {\bibfnamefont {R.}~\bibnamefont
  {Pezer}}\ and\ \bibinfo {author} {\bibfnamefont {H.}~\bibnamefont {Buljan}},\
  }\bibfield  {title} {\bibinfo {title} {{Momentum distribution dynamics of a
  Tonks-Girardeau gas: Bragg reflections of a quantum many-body wave packet}},\
  }\href {https://doi.org/10.1103/PhysRevLett.98.240403} {\bibfield  {journal}
  {\bibinfo  {journal} {Phys. Rev. Lett.}\ }\textbf {\bibinfo {volume} {98}},\
  \bibinfo {pages} {240403} (\bibinfo {year} {2007})}\BibitemShut {NoStop}%
\bibitem [{\citenamefont {Gaunt}\ \emph {et~al.}(2013)\citenamefont {Gaunt},
  \citenamefont {Schmidutz}, \citenamefont {Gotlibovych}, \citenamefont
  {Smith},\ and\ \citenamefont {Hadzibabic}}]{Gaunt2013}%
  \BibitemOpen
  \bibfield  {author} {\bibinfo {author} {\bibfnamefont {A.~L.}\ \bibnamefont
  {Gaunt}}, \bibinfo {author} {\bibfnamefont {T.~F.}\ \bibnamefont
  {Schmidutz}}, \bibinfo {author} {\bibfnamefont {I.}~\bibnamefont
  {Gotlibovych}}, \bibinfo {author} {\bibfnamefont {R.~P.}\ \bibnamefont
  {Smith}},\ and\ \bibinfo {author} {\bibfnamefont {Z.}~\bibnamefont
  {Hadzibabic}},\ }\bibfield  {title} {\bibinfo {title} {Bose-einstein
  condensation of atoms in a uniform potential},\ }\href
  {https://doi.org/10.1103/PhysRevLett.110.200406} {\bibfield  {journal}
  {\bibinfo  {journal} {Phys. Rev. Lett.}\ }\textbf {\bibinfo {volume} {110}},\
  \bibinfo {pages} {200406} (\bibinfo {year} {2013})}\BibitemShut {NoStop}%
\bibitem [{\citenamefont {Kurtsiefer}\ \emph {et~al.}(1997)\citenamefont
  {Kurtsiefer}, \citenamefont {Pfau},\ and\ \citenamefont
  {Mlynek}}]{Kurtsiefer1997}%
  \BibitemOpen
  \bibfield  {author} {\bibinfo {author} {\bibfnamefont {C.}~\bibnamefont
  {Kurtsiefer}}, \bibinfo {author} {\bibfnamefont {T.}~\bibnamefont {Pfau}},\
  and\ \bibinfo {author} {\bibfnamefont {J.}~\bibnamefont {Mlynek}},\
  }\bibfield  {title} {\bibinfo {title} {{Measurement of the wigner function of
  an ensemble of helium atoms}},\ }\href {https://doi.org/10.1038/386150a0}
  {\bibfield  {journal} {\bibinfo  {journal} {Nature}\ }\textbf {\bibinfo
  {volume} {386}},\ \bibinfo {pages} {150} (\bibinfo {year}
  {1997})}\BibitemShut {NoStop}%
\bibitem [{\citenamefont {Hornberger}\ \emph {et~al.}(2003)\citenamefont
  {Hornberger}, \citenamefont {Uttenthaler}, \citenamefont {Brezger},
  \citenamefont {Hackerm{\"{u}}ller}, \citenamefont {Arndt},\ and\
  \citenamefont {Zeilinger}}]{Hornberger2003}%
  \BibitemOpen
  \bibfield  {author} {\bibinfo {author} {\bibfnamefont {K.}~\bibnamefont
  {Hornberger}}, \bibinfo {author} {\bibfnamefont {S.}~\bibnamefont
  {Uttenthaler}}, \bibinfo {author} {\bibfnamefont {B.}~\bibnamefont
  {Brezger}}, \bibinfo {author} {\bibfnamefont {L.}~\bibnamefont
  {Hackerm{\"{u}}ller}}, \bibinfo {author} {\bibfnamefont {M.}~\bibnamefont
  {Arndt}},\ and\ \bibinfo {author} {\bibfnamefont {A.}~\bibnamefont
  {Zeilinger}},\ }\bibfield  {title} {\bibinfo {title} {{Collisional
  Decoherence Observed in Matter Wave Interferometry}},\ }\href
  {https://doi.org/10.1103/PhysRevLett.90.160401} {\bibfield  {journal}
  {\bibinfo  {journal} {Phys. Rev. Lett.}\ }\textbf {\bibinfo {volume} {90}},\
  \bibinfo {pages} {4} (\bibinfo {year} {2003})}\BibitemShut {NoStop}%
\bibitem [{\citenamefont {Hornberger}\ \emph {et~al.}(2012)\citenamefont
  {Hornberger}, \citenamefont {Gerlich}, \citenamefont {Haslinger},
  \citenamefont {Nimmrichter},\ and\ \citenamefont {Arndt}}]{Hornberger2012a}%
  \BibitemOpen
  \bibfield  {author} {\bibinfo {author} {\bibfnamefont {K.}~\bibnamefont
  {Hornberger}}, \bibinfo {author} {\bibfnamefont {S.}~\bibnamefont {Gerlich}},
  \bibinfo {author} {\bibfnamefont {P.}~\bibnamefont {Haslinger}}, \bibinfo
  {author} {\bibfnamefont {S.}~\bibnamefont {Nimmrichter}},\ and\ \bibinfo
  {author} {\bibfnamefont {M.}~\bibnamefont {Arndt}},\ }\bibfield  {title}
  {\bibinfo {title} {{Colloquium: Quantum interference of clusters and
  molecules}},\ }\href {https://doi.org/10.1103/RevModPhys.84.157} {\bibfield
  {journal} {\bibinfo  {journal} {Rev. Mod. Phys.}\ }\textbf {\bibinfo {volume}
  {84}},\ \bibinfo {pages} {157} (\bibinfo {year} {2012})}\BibitemShut
  {NoStop}%
\bibitem [{\citenamefont {Chapman}\ \emph
  {et~al.}(1995{\natexlab{b}})\citenamefont {Chapman}, \citenamefont {Hammond},
  \citenamefont {Lenef}, \citenamefont {Schmiedmayer}, \citenamefont
  {Rubenstein}, \citenamefont {Smith},\ and\ \citenamefont
  {Pritchard}}]{Chapman1995a}%
  \BibitemOpen
  \bibfield  {author} {\bibinfo {author} {\bibfnamefont {M.~S.}\ \bibnamefont
  {Chapman}}, \bibinfo {author} {\bibfnamefont {T.~D.}\ \bibnamefont
  {Hammond}}, \bibinfo {author} {\bibfnamefont {A.}~\bibnamefont {Lenef}},
  \bibinfo {author} {\bibfnamefont {J.}~\bibnamefont {Schmiedmayer}}, \bibinfo
  {author} {\bibfnamefont {R.~A.}\ \bibnamefont {Rubenstein}}, \bibinfo
  {author} {\bibfnamefont {E.}~\bibnamefont {Smith}},\ and\ \bibinfo {author}
  {\bibfnamefont {D.~E.}\ \bibnamefont {Pritchard}},\ }\bibfield  {title}
  {\bibinfo {title} {{Photon scattering from atoms in an atom interferometer:
  Coherence lost and regained}},\ }\href
  {https://doi.org/10.1103/PhysRevLett.75.3783} {\bibfield  {journal} {\bibinfo
   {journal} {Phys. Rev. Lett.}\ }\textbf {\bibinfo {volume} {75}},\ \bibinfo
  {pages} {3783} (\bibinfo {year} {1995}{\natexlab{b}})}\BibitemShut {NoStop}%
\bibitem [{\citenamefont {Sonnentag}\ and\ \citenamefont
  {Hasselbach}(2007)}]{Sonnentag2007a}%
  \BibitemOpen
  \bibfield  {author} {\bibinfo {author} {\bibfnamefont {P.}~\bibnamefont
  {Sonnentag}}\ and\ \bibinfo {author} {\bibfnamefont {F.}~\bibnamefont
  {Hasselbach}},\ }\bibfield  {title} {\bibinfo {title} {{Measurement of
  decoherence of electron waves and visualization of the quantum-classical
  transition}},\ }\href {https://doi.org/10.1103/PhysRevLett.98.200402}
  {\bibfield  {journal} {\bibinfo  {journal} {Phys. Rev. Lett.}\ }\textbf
  {\bibinfo {volume} {98}},\ \bibinfo {pages} {200402} (\bibinfo {year}
  {2007})}\BibitemShut {NoStop}%
\end{thebibliography}%

\end{document}